\magnification=1200
\parskip=10pt plus 5pt
\parindent=12pt
\baselineskip=18pt
\input amssym.def
\input amssym
\input epsf.tex
\line{\hfil{DAMTP-R/96/8~}}
\line{\hfil{hep-th/9702070~~~}}
\pageno=0
\footline={\ifnum \pageno < 1 \else \hss \folio \hss \fi }
\vskip 5pt
\centerline{\bf SCALAR FIELD THEORY IN CURVED SPACE}
\centerline{\bf AND THE DEFINITION OF MOMENTUM}
\vskip .05in
\centerline{Simon Davis}
\vskip .05in
\centerline{Department of Applied Mathematics and Theoretical Physics}
\vskip 1pt
\centerline{University of Cambridge}
\vskip 1pt
\centerline{Silver Street, Cambridge CB3 9EW}
\vskip 1pt
\centerline{and}
\vskip 1pt
\centerline{School of Mathematics and Statistics}
\vskip 1pt
\centerline{University of Sydney}
\vskip 1pt
\centerline{NSW 2006, Australia}
\vskip 1pt
\noindent
{\bf Abstract.}  Some general remarks are made about the quantum theory
of scalar fields and the definition of momentum in curved space.  Special
emphasis is given to field theory in anti-de Sitter space, as
it represents a maximally symmetric space-time of constant curvature
which could arise in the local description of matter interactions in the
small regions of space-time.  Transform space rules for evaluating Feynman 
diagrams in Euclidean anti-de Sitter space are initially defined using 
eigenfunctions based on generalized plane waves.  It is shown that, for a 
general curved space, the rules associated with the vertex are dependent 
on the type of interaction being considered.  A condition for eliminating 
this dependence is given.  It is demonstrated that the vacuum and propagator 
in conformally flat coordinates in anti-de Sitter space are equivalent to 
those analytically continued from $H^4$ and that transform space rules based 
on these coordinates can be used more readily.  A proof of the analogue of
Goldstone's theorem in anti-de Sitter space is given using a generalized
plane wave representation of the commutator of the current and the scalar
field.  It is shown that the introduction of curvature in the space-time
shifts the momentum by an amount which is determined by the Riemann tensor
to first order, and it follows that there is a shift in both the momentum
and mass scale in anti-de Sitter space.   

\vfill
\eject

\noindent
{\bf 1. Introduction}

Quantum field theory in curved space has been studied for more than twenty
years as a preliminary step towards developing a theory of quantum gravity.  
While it has been assumed that calculations in a classical space-time do
not represent a complete quantum theory of matter interacting with the
gravitational field, they could be useful in describing the dynamics of 
particle scattering, as the energy-momentum of any point-particle field  
induces curvature in the space-time background.  
Recent considerations in unification of the elementary particle
interactions and superstring effective actions suggest that the 
standard model can be derived from a ten-dimensional theory based    
on the spinor space 
$T~=~{\Bbb R} \otimes {\Bbb C} \otimes {\Bbb H} \otimes {\Bbb O}$ and the 
Clifford algebra $R_{1,9}$ [1][2] and that the higher-order curvature terms
representing corrections to the standard Einstein-Hilbert action can
arranged in a super-Yang-Mills theory with bosonic gauge group SO(1,9), 
with the torsion in the connection given by the field strength of the 
anti-symmetric tensor field [3].  At short distances, the gravitational
force might then also be described by the exchange of quanta in this theory.  
Consequently, it is consistent to consider matter 
interactions on a background which is initially selected to be flat but then 
is curved by energy-momenta associated with the gravitational and matter 
quanta.

If it is assumed, in particular, that the energy-momenta of the quantum
fields curve the space-time and still maintain maximal symmetry, the
local geometry will resemble a manifold such as anti-de Sitter space,
while the global geometry would remain essentially flat.
Quantum field theory in anti-de Sitter space represents one of the most 
tractable quantum field theories on a curved manifold, and it has also arisen
in the study of gauged supergravity.  The techniques of Minkowski space-time 
quantum field theory shall be adapted to anti-de Sitter space to 
define appropriate momentum variables and construct Feynman rules, obtain 
the Lehmann spectral representation of the two-point function, prove the 
analogue of Goldstone's theorem and investigate the relation between 
field-theoretic calculations in a local region of space-time with 
curved geometry and string scattering in a target space with a 
specified global geometry.

It is shown in this paper how configuration and transform [momentum] space 
rules can be written for scalar field theory in anti-de Sitter space.  In the 
configuration space rules, factors 
are obtained for each vertex, propagator, incoming particle and outgoing 
particle and a symmetry factor may be assigned for each Feynman diagram.
Similar rules can be formulated for every vertex, propagator and loop transform
variable in transform space, and again, a symmetry factor may be associated 
with each diagram.  The one-loop box diagram for $\lambda\phi^3 /3!$ theory is
evaluated using these rules.

A general feature of the adaptation of the momentum space rules to curved space
is that it is necessary to assign factors associated with each vertex that are
dependent on the type of interaction being considered.  Independence of the
vertex rules with respect to the type of interaction becomes a constraint
on the eigenfunctions representing the incoming and outgoing particles. 
In anti-de Sitter space, it is demonstrated that the maximal symmetry of the 
space allows for a choice of coordinates in which these constraints may be 
simplified. These methods are compared to other perturbation theory techniques
and comments are made about the renormalizability of the theory.

\noindent
{\bf 2. Configuration and Dual Space Rules for Scalar Theory in Euclidean}
\hfil\break
{\bf {\phantom {....}} Anti-de Sitter Space}

Euclidean anti-de Sitter space is the hyperboloid $H^4$ 
$$-x_0^2-x_1^2-x_2^2-x_3^2+x_4^2~=~1
\eqno(1)$$
embedded in the five-dimensional pseudo-Euclidean space with metric
$$ds^2~=~dx_0^2+dx_1^2+dx_2^2+dx_3^2-dx_4^2
\eqno(2)$$
and it can be represented as the coset space SO(4,1)/SO(4).  Using the 
hyperbolic and spherical angles $\theta_1$, ..., $\theta_4$, defined by
$$\eqalign{x_0~&=~sinh~\theta_4~sin~\theta_3~sin~\theta_2~sin~\theta_1
\cr
x_1~&=~sinh~\theta_4~sin~\theta_3~sin~\theta_2~cos~\theta_1
\cr
x_2~&=~sinh~\theta_4~sin~\theta_3~cos~\theta_2
\cr
x_3~&=~sinh~\theta_4~cos~\theta_3
\cr
x_4~&=~cosh~\theta_4
\cr
0 \le \theta_1 < 2\pi,&~0\le \theta_2,~ \theta_3 < \pi,~0\le 
\theta_4 < \infty 
\cr}
\eqno(3)$$
the Laplacian in $H^4$ is
$$\eqalign{\nabla^\mu \nabla_\mu = {1\over {sinh^3 \theta_4}}&
{\partial\over {\partial\theta_4}}
sinh^3 \theta_4 {\partial\over {\partial \theta_4}}+
{1\over {sinh^2 \theta_4 sin^2 \theta_3}} 
{\partial\over {\partial\theta_3}} sin^2 \theta_3 {\partial
\over {\partial \theta_3}}
\cr
&+{1\over {sinh^2\theta_4 sin^2\theta_3 sin \theta_2}}
{\partial\over {\partial \theta_2}}~sin \theta_2 ~{\partial\over 
{\partial\theta_2}}
+{1\over {sinh^2 \theta_4 sin^2 \theta_3 sin^2 \theta_2}} 
{{\partial^2}\over {\partial \theta_1^2}} 
\cr}
\eqno(4)$$
which has the following basis of eigenfunctions [4]
$$\eqalign{\Theta_K^\sigma (x) &~=~ B_K^\sigma (sinh~\theta_4)^{-1} 
{\cal P}_{\sigma+1}^{-k_0-1} (cosh~\theta_4)
\cr 
&~~~~~~\cdot \prod_{j=0,1} C_{k_j-k_{j+1}}^{1-{j\over 2}+k_{j+1}} 
(cos~\theta_{3-j}^4)
sin^{k_{j+1}}\theta_{3-j} e^{i k_2 \theta_1}
\cr
B_K^\sigma &~=~ {{2 (-1)^{k_0} \Gamma(\sigma-1)}\over {\Gamma(\sigma-k_0+1)}}
\cr
&~~~~\left[{\Gamma(2)}^{-1} \prod_{j=0,1} 
{{2^{2k_{j+1}-j}(k_j-k_{j+1})!}\over {\sqrt\pi}{\Gamma(k_j+k_{j+1}+2-j)}}
\cdot (2-j+2k_j) 
\Gamma^2(1-{j\over 2}+k_{j+1})\right]^{1\over 2}
\cr
K &~=~ (k_0,k_1,k_2)
\cr}
\eqno(5)$$
where ${\cal P}_l^m(z)$ and $C_m^p(t)$ are the associated Legendre functions
and Gegenbauer polynomials respectively.
From the relations
$${\sqrt {z^2-1}} {{d {\cal P}_l^m (z)}\over {dz}}~-~
{{mz}\over {\sqrt {z^2-1}}}
{\cal P}_l^m(z)~=~{\cal P}_l^{m+1}(z)
\eqno(6)$$
it follows that the corresponding eigenvalues are 
$(k-1)(k+2)+(\sigma-k+1)(\sigma+k+2)=\sigma(\sigma+3)$.
The normalization factor associated with the eigenfunctions 
$\Theta_K^\sigma(x)$
determined by
$${1\over {(N(\sigma,k))^2}}~
\int_{H^4}~dx~\Theta_K^\sigma(x){\bar \Theta}_K^{\sigma}(x) 
~=~1
\eqno(7)$$
is
$$N(\sigma, k)~=~ \left[{{(-1)^{-1-k}}\over {\pi^2}}(\sigma+{1\over 2})
{{\Gamma^2(\sigma-1)}\over {\Gamma(\sigma-k+1)
\Gamma(\sigma+k+3)}}\right]^{1\over 2},~~~ k\equiv k_0
\eqno(8)$$
The configuration space rules for scalar field theory can then be given:
\hfil\break
1.  To each vertex attach a factor $-i \lambda$ and $\int d^4x~{\sqrt{g(x)}}$.
\hfil\break
2.  A propagator $i \Delta_F(x, x^\prime)$ for each line from $x$ to 
$x^\prime$.
\hfil\break
3.  A factor of ${{\Theta_K^\sigma(x)}\over {N(\sigma, k)}}$ for an incoming
particle with transform numbers $\sigma$, K.
\hfil\break
4.  A factor of ${{\Theta_{\bar K}^{-\sigma-3}(x)}\over {N(\sigma,k)}}$ for an 
outgoing particle with transform space numbers $-\sigma-3$, 
\hfil\break
{\phantom {....}} ${\bar K}$.
\hfil\break
5.  A symmetry factor ${1\over g}$ where $g$ is the order of the symmetry group
of the diagram for 
\hfil\break
{\phantom {....}}
operations that leave the external lines fixed.

An integral transform may be used to generalize momentum-space Feynman rules
to $H^4$.  If the coefficients $a_K(\sigma)$ are defined to be
$\int_{H^4} dx f(x) \Theta_K^\sigma(x) $, then
$$f(x)~=~{i\over {16 \pi^2 \Gamma(2)}} \sum_K 
\int_{(Re~\sigma_0)-i\infty}^{(Re~\sigma_0)+i\infty}~
d\sigma{{\Gamma(\sigma+3)}\over {\Gamma(\sigma)}}
ctg~\pi\sigma~a_K(\sigma) \Theta_{\bar K}^{-\sigma-3}(x)~~~~~~-3<Re~\sigma_0<1
\eqno(9)$$
which follows from the delta-function identity
$${i\over {16 \pi^2 \Gamma(2)}} \sum_K 
\int_{(Re~\sigma_0)-i \infty}^{(Re~\sigma_0)+i\infty}
d\sigma {{\Gamma(\sigma+3)}\over {\Gamma(\sigma)}} \Theta_K^\sigma(x)
\Theta_{\bar K}^{-\sigma-3} (x^\prime)~=~ \delta(x-x^\prime)
\eqno(10)$$
Then the Feynman propagator may expressed as
$$\Delta_F(x,x^\prime)~=~{i\over {16 \pi^2 \Gamma(2)}} 
\sum_K \int_{(Re~\sigma_0)-i \infty}^{(Re~\sigma_0)+i \infty} 
d\sigma {{\Gamma(\sigma+3)}\over {\Gamma(\sigma)}} 
ctg~\pi\sigma \Delta_F(k,\sigma) \Theta_K^\sigma(x) 
\Theta_{\bar K}^{-\sigma-3}(x^\prime)
\eqno(11)$$
if
$$\Delta_F(K,\sigma)~=~{1\over {\sigma(\sigma+3)+m^2}}
\eqno(12)$$
where $m$ is the mass of the scalar field. 
The transform space, or dual space, rules for scalar field theory would be
\hfil\break
1.  A factor $-i\lambda$ and $f_{\{K\}}^{\{\sigma\}}$ for each vertex.
\hfil\break
2.  ${i\over {\sigma(\sigma+3)+m^2}}$ for each propagator.
\hfil\break
3.  ${i\over {16 \pi^2 \Gamma(2)}} \sum_K 
\int_{Re~\sigma_0-i\infty}^{Re~\sigma_0+i\infty}~
d\sigma {{\Gamma(\sigma+3)}\over {\Gamma(\sigma)}}~ ctg~\pi\sigma$ for each
loop transform variable.
\hfil\break
4.  A factor ${1\over g}$ where $g$ is the order of the symmetry group of the
diagram.

\noindent
{\bf 3. The $\lambda \phi^3/3!$ Box Diagram in $H^4$ and Vertex Factors}

The factor $f_{\{K\}}^{\{\sigma\}}$, which is necessary for equivalence 
between the integrals derived from configuration space and dual 
space rules, depends on the type of interaction being considered.  
For example, the box diagram (Fig. 1) in $\lambda\phi^3/3!$ theory can 
be evaluated using either set of rules.  Configuration space rules give
$$\eqalign{{{(-i\lambda)^4}\over 4} \int~d^4x_1~d^4x_2~d^4x_3~d^4x_4~&~ 
{\sqrt{g(x_1)}}~{\sqrt{g(x_2)}}~{\sqrt{g(x_3)}}~{\sqrt{g(x_3)}}~ 
\cr
\cdot ~(i\Delta_F(x_1,x_2))&(i\Delta_F(x_2,x_3))(i\Delta_F(x_3,x_4))
(i\Delta(x_4,x_1))
\cr
~\cdot{{\Theta_{K_1}^{\sigma_1}(x_1)}\over {N(\sigma_1, K_1)}}& 
{{\Theta_{K_2}^{\sigma_2}(x_2)}\over {N(\sigma_2, K_2)}} 
{{\Theta_{{\bar K}_3}^{-\sigma_3-3}(x_3)}\over {N(\sigma_3, K_3)}} 
{{\Theta_{{\bar K}_4}^{-\sigma_4-3}(x_4)}\over {N(\sigma_4, K_4)}}
\cr}
\eqno(13)$$

\vbox{
\epsfysize=2in
\epsfxsize=3in
\centerline{\epsfbox{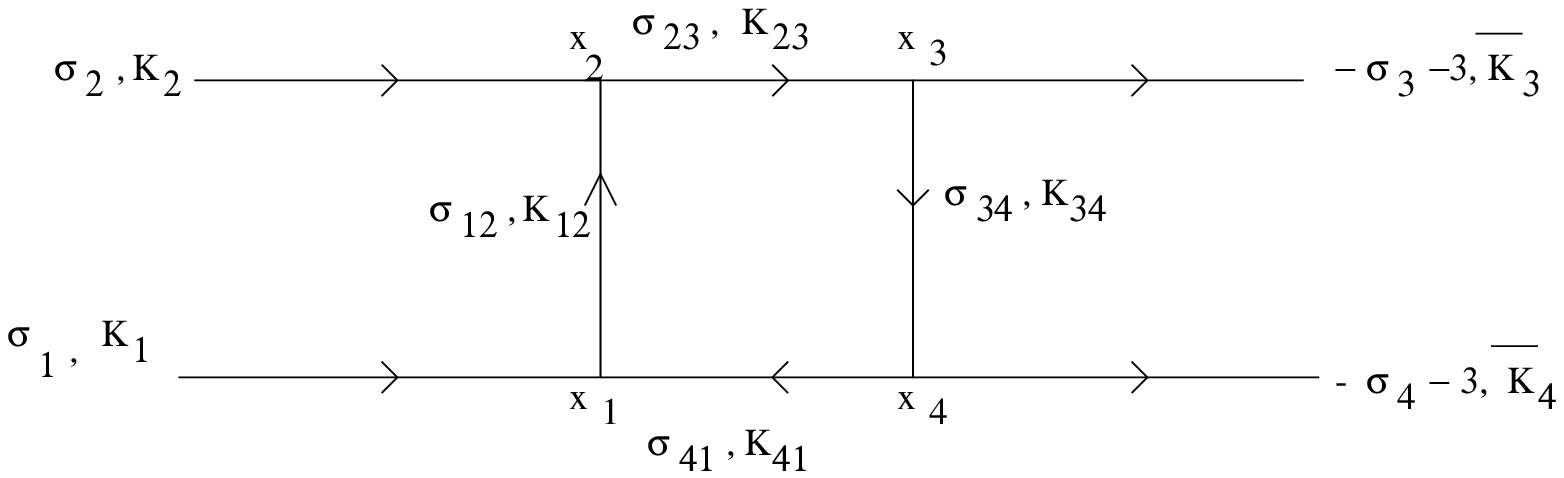}}
\vskip 0.2in
\noindent{\bf Fig. 1. The $\lambda{{ \phi^3}\over {3!}}$ box diagram
in Euclidean anti-de Sitter space.}}

\noindent
while transform space rules give
$$\eqalign{{{(-i\lambda)^4}\over 4}
\left({1\over {16 \pi^2}\Gamma(2)}\right)^4
\sum_{{K_{12},K_{23}}\atop {K_{34},K_{41}}} 
\int_{(Re~\sigma_0)-i\infty}^{(Re~\sigma_0)+i\infty}
d&\sigma_{12} d\sigma_{23} 
d\sigma_{34} d\sigma_{41} {{\Gamma(\sigma_{12}+3)}\over {\Gamma(\sigma_{12})}}
~ctg~\pi\sigma_{12}
\cr
{{\Gamma(\sigma_{23}+3)}\over {\Gamma(\sigma_{23})}} 
&ctg~\pi \sigma_{23} 
{{\Gamma(\sigma_{34}+3)}\over {\Gamma(\sigma_{34})}}~
ctg~\pi\sigma_{34} 
{{\Gamma(\sigma_{41}+3)}\over {\Gamma(\sigma_{41})}}
\cr
ctg~\pi\sigma_{41}\cdot &(i\Delta_F(\sigma_{12}))
(i\Delta_F(\sigma_{23}))(i\Delta_F(\sigma_{34}))(i\Delta_F(\sigma_{41}))
\cr
&~\cdot\prod_{vert.}~f_{\{K\}}^{\{\sigma\}}(vert.)
\cr}
\eqno(14)$$

Applying the formula (11) to equation (13) implies that
$$\eqalign{\prod_{vert.}~f_{\{K\}}^{\{\sigma\}}(vert.)
~=~\int d^4x_1~{\sqrt{g(x_1)}}~&\Theta_{\bar K_{12}}^{-\sigma_{12}-3}(x_1) 
~\Theta_{K_{41}}^{\sigma_{41}}(x_1)~{{\Theta_{K_1}^{\sigma_1}(x_1)}
\over {N(\sigma_1,K_1)}}
\cr
\int~d^4x_2~&{\sqrt{g(x_2)}}~\Theta_{K_{12}}^{\sigma_{12}}(x_2) 
~\Theta_{{\bar K}_{23}}^{-\sigma_{23}-3}(x_2)
{{~\Theta_{K_2}^{\sigma_2}(x_2)}\over {N(\sigma_2, K_2)}}
\cr
\int~d^4x_3~&{\sqrt{g(x_3)}}~\Theta_{K_{23}}^{\sigma_{23}}(x_3) 
~\Theta_{{\bar K}_{34}}^{-\sigma_{34}-3}(x_3) 
~{{\Theta_{{\bar K}_3}^{-\sigma_3-3}(x_3)}
\over {N(\sigma_3, K_3)}}
\cr
\int~d^4x_4~&{\sqrt{g(x_4)}} ~\Theta_{K_{34}}^{\sigma_{34}}(x_4) 
~\Theta_{{\bar K}_{41}}^{-\sigma_{41}-3}(x_4) 
~{{\Theta_{{\bar K}_4}^{-\sigma_4-3}(x_4)}
\over {N(\sigma_4, K_4)}}
\cr}
\eqno(15)$$

In Euclidean space, this product would be
$$\eqalign{\int~d^4x_1~&e^{-ip_{12}\cdot x_1} e^{ip_{41}\cdot x_1} 
e^{ip_1\cdot x_1}
\int~d^4x_2~e^{ip_{12}\cdot x_2} e^{-ip_{23}\cdot x_2} e^{ip_2\cdot x_2}
\cr
&\int~d^4x_3~e^{ip_{23}\cdot x_3} e^{-ip_{34}\cdot x_3} e^{-ip_3\cdot x_3}
\int~d^4x_4~e^{ip_{34}\cdot x_4} e^{-ip_{41}\cdot x_4} e^{-ip_4\cdot x_4}
\cr
~&=~\delta(p_{12}-p_1-p_{41})~\delta(p_2+p_{12}-p_{23})~\delta(p_3+p_{34}-
p_{23})
~\delta(p_4+p_{41}-p_{34})
\cr}
\eqno(16)$$
which ensures momentum conservation at each vertex and for the external 
states in the scattering process.

The appearance of the factors $f_{\{K\}}^{\{\sigma\}}$ for each vertex in the 
transform space rules is a feature of quantum field theory in a 
curved space.  They arise even for the hyperboloid $H^4$, which has maximal 
symmetry.  These factors depend on the type of interaction being considered 
because the number of eigenfunctions in each integral defining 
$f_{\{K\}}^{\{\sigma\}}$  is equal to the number of lines at each vertex in 
the Feynman diagram.  For a general curved Riemannian space, an integral 
involving three or more eigenfunctions of the Laplacian would not reduce to a 
delta-function containing the transform space variables, whereas, in Euclidean
space, the reduction, which is independent of the type of interaction, does 
occur and only implies the physically necessary constraint of momentum
conservation.  The condition that must be obeyed by the eigenfunctions
to obtain a delta-function is that they form a group under pointwise 
multiplication, as the number of eigenfunctions in the integral then can be 
reduced until the orthogonality relation (7) may be used.  As this condition
is satisfied by $e^{ip\cdot x}$, it is useful to consider generalized plane 
waves in curved spaces, and, specifically for $H^4$, to clarify their 
connection with the eigenfunctions (5).

\noindent
{\bf 4. Generalized Plane Waves and the Fourier Transform on $H^4$}

The maximal symmetry of $H^4$ allows for the construction of generalized plane
waves and the Fourier transform using group theoretic methods.  Horocycles, 
the analogue of hyperplanes, may be defined by the action of a nilpotent 
subgroup in the Iwasawa decomposition of SO(4,1).  Generalized plane waves are 
represented by eigenfunctions of the Laplacian operator that are exponential 
functions of the hyperbolic distance from a chosen origin to the horocycle.

Specifically, the Iwasawa decomposition of a semisimple Lie algebra is
${\cal G}~=~{\cal H}~\oplus~{\cal A}~\oplus~{\cal N}$ where ${\cal H}$ is the 
maximal compact subalgebra, ${\cal A}$ is a maximal abelian subalgebra in the
space spanned by the non-compact generators and ${\cal N}$ is a nilpotent 
subalgebra defined the positive roots of ${\cal G}$.  Given a point $o$ chosen 
to be the origin in a Cartan symmetric space $G/H$, the fundamental 
horocycle is $\xi_0=N\cdot o$.  All other horocycles may be expressed as
$ha\cdot \xi_0$, $h\in H, ~a \in A$.  This representation is not 
unique; defining $M$ to be the centralizer of ${\cal A}$ in $H$, $h$ and 
$h^\prime$ give the same horocycles if they belong to the same coset in H/M, 
which is also known as the boundary of the symmetric space [5].      

The Fourier transform is given by
$${\tilde f(\lambda,b)}~=~\int_{G/H}~dx~f(x)~e^{-(i\lambda+\rho)(r(x,b))} 
\eqno(17)$$
where $\rho~=~{1\over 2} \sum_{\alpha>0}~dim~{\cal G_\alpha}$, with 
${\cal G_\alpha}$ being the vector space spanned by generators in ${\cal G}$ 
associated with the positive root $\alpha$, and $r(x,b)$ is the distance 
from the
origin $o$ to the horocycle passing through $x$ with normal $b$.  The  inverse 
transform is
$$f(x)~=~ \int_{{\cal A}^\ast} \int_{H/M}~d\lambda~db~{\tilde f}(\lambda,b) 
~e^{(i\lambda-\rho)(r(x,b))}~\vert {\hat c}(\lambda)\vert^{-2}  
\eqno(18)$$
where ${\cal A^\ast}$ is the space of functionals on ${\cal A}$ and 
${\hat c}(\lambda)$ is a spectral function selecting only those values of 
$\lambda$ corresponding
to irreducible representations of $G$ [6].  In the standard treatise on the
Fourier transform on symmetric spaces [7], the exponent $\rho$ appears
with a positive sign.  However, the negative sign is also appropriate for 
$H^4$, considering the eigenvalues of the d'Alembertian, mentioned immediately
following equation (6), and this is confirmed by the integral transform
for $H^{n-1}$, where the contour of integration is 
$-{{n-2}\over 2}+i \lambda$ [4].  The difference between the two transforms
can be absorbed into the spectral function. 

The action of the nilpotent subgroup in the Iwasawa decomposition of SO(4,1)
on the chosen origin $o=(0,0,0,0,1)$ defines the fundamental horocycle on 
$H^4$ which is the intersection of a hyperboloid with a hyperplane
$$\eqalign{-x_0^2-x_1^2-x_2^2-x_3^2+x_4^2 &~=~ 1
\cr
-x_0+x_4&~=~1
\cr}
\eqno(19)$$

\vbox{
\epsfysize=2.2in
\centerline{\epsfbox{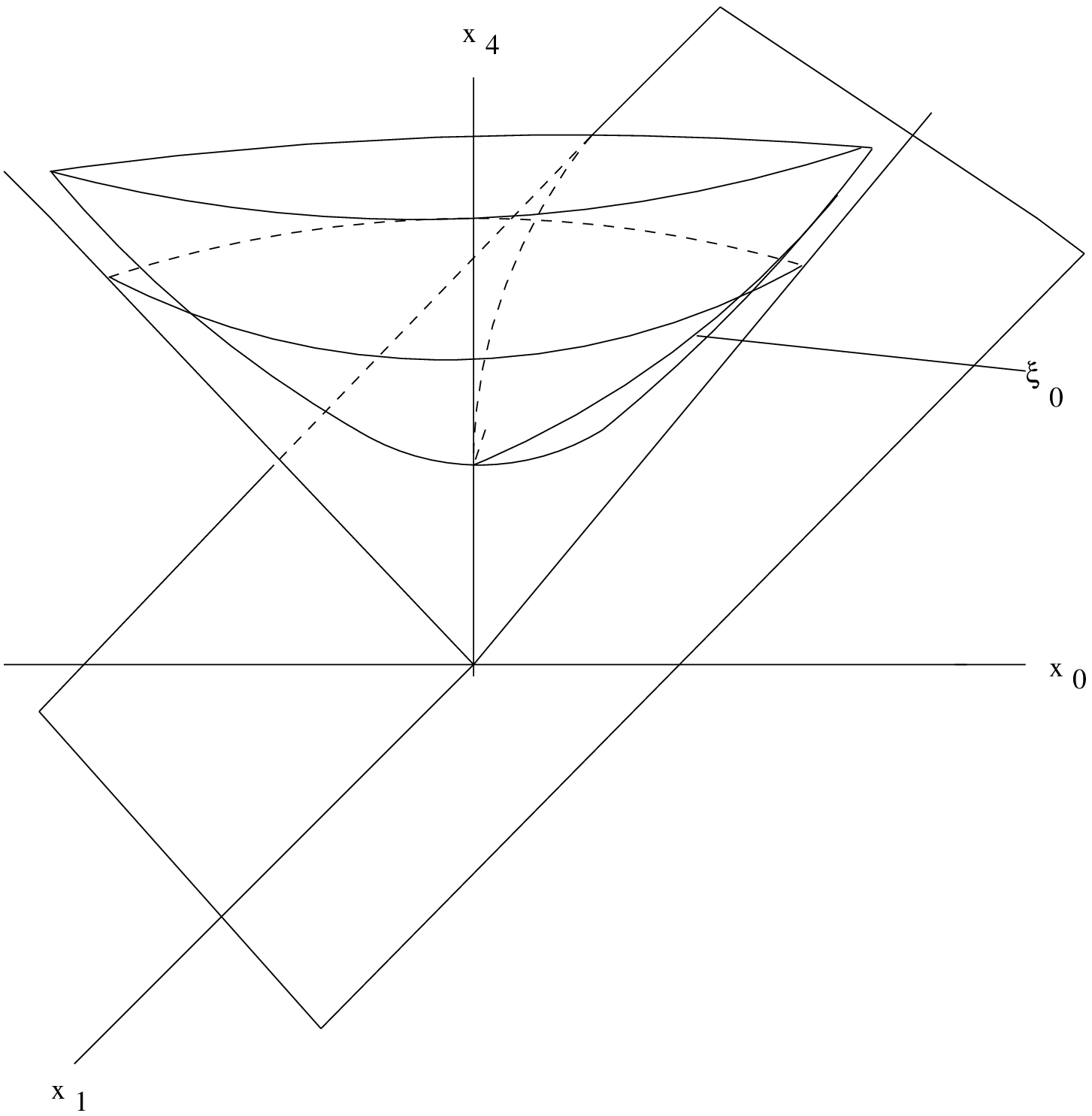}}
\vskip 0.2in
\noindent{\bf Fig. 2.  The fundamental horocycle $\xi_0$.}}

It can be shown, either group theoretically or geometrically [6], that the 
distance from the origin to the horocycle passing through $x$ with normal 
$b$ is $ln [x, h\xi_0]$, where $hM$ is the coset corresponding
to $b$, $\xi_0$ represents the null vector (1,0,0,0,1) on 
the cone
$-\xi_0^2-\xi_1^2-\xi_2^2-\xi_3^2+\xi_4^2=0$ and $[x,\xi]$ is the
scalar product $-x_0\xi_0-x_1\xi_1-x_2\xi_2-x_3\xi_3+x_4\xi_4$.  
The calculation of the hyperbolic distance is given in the appendix. 
Since
$\rho={3\over 2}$, the generalized plane waves on $H^4$ will be 
$[x, \xi]^{-i\lambda-{3\over 2}}$, which are eigenfunctions of the Laplacian
with eigenvalue $-(\lambda^2+{9\over 4})$, and the Fourier transform will be
$$f(\eta, \lambda)~=~\int_{H^4}~dx~f(x)~[x, \xi]^{-i\lambda-{3\over 2}} 
\eqno(20)$$  
Since the null vectors on the cone are in one-to-one correspondence with 
points on $S^3$, these functions can be expanded in terms of basis 
eigenfunctions on the sphere
$$\eqalign{{\tilde f}(\xi, \lambda)~&=~\sum_K~a_K(\lambda)~\Xi_K(\xi)
\cr
a_K(\lambda)~&=~\int_{S^3}~d\eta~\tilde f(\xi,\lambda)~{\bar \Xi}_K(\eta)
\cr}
\eqno(21)$$
The equivalence of the Fourier coefficients $a_K(\lambda)$ in equation (21)
and $a_K(\sigma)$ defined just before equation (9) follows from an identity
containing the associated Legendre functions [4]
$$\Theta_K^\sigma(x)~=~\int_{S^3}~d\xi~[x,\xi]^\sigma~{\bar \Xi}_K(\eta)
\eqno(22)$$
This relation reveals the plane-wave content in the configuration space 
and dual space rules based on the eigenfunctions $\Theta_K^\sigma(x)$.
One may note further that the propagator for a massive scalar field [5][8] 
may also be written as
$$\Delta_F(\lambda)~=~{1\over {a^2\lambda^2~-~m^2~+~{9\over 4}a^2}}
\eqno(23)$$
so that the variable $\lambda$ could be interpreted as the norm of the
four-momentum with a shift in the zero-point of the squared-mass scale by
$-{9\over 4}a^2$, where
$a^{-1}$ is the radius of curvature of the hyperboloid, which is customarily 
set equal to 1.  Analytic continuation to anti-de Sitter space is achieved
by replacing $\lambda$ by $-i\lambda$ and $a$ by $ia$.

\vfill\eject
\noindent
{\bf 5. Spectral Representation of the Propagator}

In flat-space field theory, non-perturbative results about S-matrix elements
can be obtained when the fields satisfy properties such as
\vskip 5pt
\noindent
${\underline{Poincare~invariance}}$
\hfil\break
$$\eqalign{
e^{i\alpha \cdot P} \phi(x) e^{-i \alpha\cdot P}~&=~\phi(x+\alpha)       
\cr
U(\Lambda) \phi(x) U(\Lambda^{-1})~&=~ \phi(\Lambda x)
\cr}
$$
\vskip 2pt
\noindent
${\underline{Locality}}$          
\hfil\break
$$[\phi(x), \phi(y)]~=~0~~~if~~(x-y)^2~<~0$$ 
\vskip 2pt
\noindent
${\underline{Spectral~assumption}}$          
\hfil\break
The eigenvalues of $P^0$, $P^2$ are non-negative.
\vskip 2pt
\noindent
${\underline{Uniqueness~of~vacuum}}$
\hfil\break          
There exists a unique vacuum $\vert 0 \rangle$
such that $P_\mu \vert 0 \rangle~=~0$, $U(\Lambda)\vert 0\rangle~=~
\vert 0 \rangle$.
\vskip 5pt
\noindent
Because of the spectral assumption, the positive frequency commutator
function can be written as 
$\Delta_+(x)~=~{-{i\over {(2\pi)^3}}}~\int~d^4k
~e^{-ik\cdot x}~\rho(k^2)\theta(k^0)$ 
where $\rho(k^2)\ge 0$ and $\rho(k^2)=0$ if $k^2\le 0$, and the Lehmann 
representation for the propagator is
$\Delta_F(p)~=~\int_0^\infty~d\sigma~{{\rho(\sigma)}
\over {p^2-\sigma+i \delta}}$.

The assumptions above can be generalized to $H^4$, with Poincare invariance
being replaced by SO(4,1) invariance for example.  However, the spectrum,
instead of starting at zero, begins at $-{9\over 4}a^2$, so that the Lehmann
representation [6][7] becomes
$$\Delta_F(p)~=~\int_{-{9\over 4}a^2}^\infty~d\sigma~{{\rho(\sigma)}
\over {p^2-\sigma+i \delta}}
\eqno(24)$$

Thus, the generalized plane wave decomposition of Green function can be used to
provide a momentum-space Lehmann representation for Feynman propagators in
anti-de Sitter space [6], leading to a non-perturbative definition of the
mass as a pole in the propagator [8].

\noindent
{\bf 6. The Vertex Factor and Products of Generalized Plane Waves}

Feynman rules may now be formulated in the configuration and dual
space using the generalized plane wave decomposition of the scalar field.
\vskip 5pt
\noindent
Configuration space rules
\hfil\break
1.  To each vertex attach a factor $-i\lambda$ and $\int~d^4x {\sqrt {g(x)}}$.
\hfil\break
2.  A propagator $i \Delta_F(x_i,x_j)$ for each line from $x_i$ to $x_j$.
\hfil\break
3.  A factor of $e^{-(i\lambda+{3\over 2})r(x,b)}$ for each particle traveling
towards $x$ with `momentum' $\lambda$ on 
\hfil\break
{\phantom{....}}
a geodesic orthogonal to the horocycle $\xi(x,b)$.
\hfil\break
4.  A factor of $e^{(i\lambda-{3\over 2})r(x,b)}$ for each particle traveling 
outwards from $x$ with `momentum' 
\hfil\break
{\phantom{....}}
$\lambda$ on a geodesic orthogonal 
to the horocycle $\xi(x,b)$.
\hfil\break
5.  Symmetry factor ${1\over g}$ where $g$ is the order of the symmetry
group of the diagram leaving 
\hfil\break
{\phantom{....}}
external lines fixed.
\vskip 5pt
\noindent
Setting $a~=~1$, the Fourier transform in $H^4$ implies the following set of
rules.
\vskip 5pt
\noindent
Dual space rules
\hfil\break
1.  A factor of $-i\lambda$ and $f^{\{\lambda\}}_{\{b\}}$ is assigned to each 
vertex.
\hfil\break
2.  A factor of ${i\over {\lambda^2-m^2+{9\over 4}}}$ for each propagator.
\hfil\break
3.  $\int_\lambda~d\lambda~\vert {\hat c}(\lambda)\vert^{-2}~\int_{S^3}~db$
for each independent loop dual space variable.
\hfil\break
4.  A symmetry factor ${1\over g}$ where $g$ is the order of the symmetry 
group of the diagram.

The usefulness of the dual space rules depends on the factor 
$\prod_{vert.}f^{\{\lambda\}}_{\{b\}}$.  The generalized plane waves have
a form similar to the plane waves $e^{ip\cdot x}$ in flat space, and
their product arises in the vertex factor.

Consider, for example, the product of two generalized plane waves
$$e^{-(i\lambda+{3\over 2})r(x,b)}
e^{-(i\lambda^\prime+{3\over 2})r(x,b^\prime)}
~=~e^{-{{[i(\lambda+\lambda^\prime)+3]}\over 2}(r(x,b)+r(x,b^\prime))}
e^{-{{i(\lambda^\prime-\lambda)}\over 2}(r(x,b^\prime)-r(x,b))}
\eqno(25)$$

The point $x$ can reached from the origin $o$ by the action of a group
element $g_x$.  Since $H$ acts transitively on the boundary $B$, there
always exists an element $h_{b^\prime}$ such that 
$b^\prime~=~g_x h_{b^\prime}\cdot b$.  It can be demonstrated that
the following addition theorem is valid [7].
$$r(x,b)~+~r(g\cdot o, g\cdot b)~=~r(g\cdot x, g\cdot b)
\eqno(26)$$
As $h_{b^\prime}\cdot o~=~o$, $x~=~g_x h_{b^\prime}\cdot o$ and
$$r(x,b)~+~r(x,b^\prime)~=~r(x,b)~+~r(g_x h_{b^\prime}\cdot o,
g_x h_{b^\prime}\cdot o)~=~r(g_x h_{b^\prime}\cdot x, g_x h_{b^\prime}\cdot b)
~=~r(g_x h_{b^\prime}\cdot x, b^\prime)
\eqno(27)
$$
As $x~=~g_x \cdot o$, $g_x H g_x^{-1}$ is the stability group of $x$.
If ${\tilde h}_{b^\prime}\in H$ is chosen appropriately,
$b^\prime~=~g_x {\tilde h}_{b^\prime} g_x^{-1}$ and
$$\eqalign{r(x,b^\prime)~-~r(x,b)
~&=~r(g_x{\tilde h}_{b^\prime}g_x^{-1}\cdot x, g_x{\tilde h}_{b^\prime}
g_x^{-1}\cdot b)~-~r(x,b)
~=~r(g_x {\tilde h}_{b^\prime} g_x^{-1}\cdot o,
 g_x{\tilde h}_{b^\prime}\cdot b)
\cr
~&=~r(g_x{\tilde h}_{b^\prime} g_x^{-1}\cdot o, b^\prime)
\cr}
\eqno(28) 
$$ 

These formulas could be used iteratively to simplify products of more than
two generalized plane wave functions, and this would be necessary for
the dual space rules to be useful in the evaluation of diagrams in
perturbation theory.   The Fourier transform (17) on $H^4$ and the
inverse transform (18) lead to the following identity
$${\tilde f}(\lambda,b)~=~\int_{{\cal A}^*}~\int_{S^3}~
d{\tilde\lambda}~d{\tilde b}~{\tilde f}({\tilde \lambda}, {\tilde b})
\vert {\hat c}({\tilde\lambda})\vert^{-2}
~\int_{H^4}~dx~e^{-(i\lambda+{3\over 2})r(x,b)}
~e^{(i{\tilde \lambda}-{3\over 2})r(x,{\tilde b})}  
\eqno(29)
$$
which implies an identity
$$\int_{H_4}~dx~e^{-(i\lambda+{3\over 2})(r(x,b))}
\cdot~e^{-(i\lambda^\prime +{3\over 2})r(x,b^\prime)}~=~
\delta(\lambda+\lambda^\prime)
~\delta(b-b^\prime) \cdot \vert {\hat c}(\lambda) \vert^2
\eqno(30)
$$
that resembles the delta function relation (16) expressing momentum 
conservation.
The concept of generalized plane waves in the hyperboloid $H^4$ and the
momentum-space
\hfil\break
Lehmann representation for the propagator in anti-de Sitter
space rely on the general theory of the Fourier transform developed for
locally symmetric spaces, for which there exist isometry-reversing
geodesics at any point [5].  Since it follows that there are spacelike
slices dividing these manifolds into two parts $M^+$ and $M^-$ and
isometries such that $\Theta(M^{\pm})~=~M^{\mp}$, it has been noted that this
is precisely the condition needed for reflection positivity [9], 
which is a property of an axiomatic quantum field theory possessing a 
positive-definite inner product in the Fock space.

\noindent
{\bf 7. Goldstone's Theorem in Anti-de Sitter Space}

These techniques can also be applied to the proof of a theorem for 
anti-de Sitter space similar to Goldstone's theorem.  The existence of 
Goldstone particles in de Sitter space has been studied previously [10].  
The proof of this theorem begins with the charges
$$Q^a~=~\int~d^3x~j^{0a}(x)~~~~~~~\partial^\mu j_\mu^a~=~0
\eqno(31)$$
satisfying commutation relations $[Q^a,Q^b]~=~if^a{}_{bc}Q^c$
generating a group $G$ and a set of scalar fields transforming under a
nontrivial representation of $G$, so that 
\hfil\break
$[Q^a, \phi_r]~=~-i T^a{}_{rs} \phi_s$.
In Minkowski space-time, the expectation value of the commutators of the 
currents associated with the original symmetry of the Lagrangian and the 
scalar fields is
$$F^a_{r\mu}(x)~=~\langle 0 \vert~[j_\mu^a(x),\phi_r]~\vert 0 \rangle
~=~\partial_\mu D(x)
\eqno(32)$$
where
$$F^a_{r\mu}(q)~=~\int~d^4x~e^{iq\cdot x}~F^a_{r\mu}(x)
~=~\int~d^4x~\partial_\mu D(x)
~=~-iq_\mu~\int~d^4x~e^{iq\cdot x}D(X)~=~-iq_\mu D(q)
\eqno(33)$$
and current conservation implies that
$$q^\mu~F^a_{r\mu}(q)~=~-iq^2~D(q)~=~0
\eqno(34)$$
and $D(q)$ contains $\delta(q^2)$, demonstrating the existence of dim $G$ - 
dim $H$ massless scalar particles where $H$ is the stability group of 
$\langle 0\vert {\underline \phi}\vert 0 \rangle~\ne~0$.  

In anti-de Sitter space, the analytic continuation to $H^4$ can be used
to obtain the Fourier representation
$$F^a_{r\mu}(\xi,\lambda)~=~\int_{H^4}~dx~\langle 0\vert~[j^a_\mu(x),\phi_r]
~\vert 0 \rangle~[x, \xi]^{-i \lambda-{3\over 2}}
\eqno(35)$$
and since $G$ acts transitively on the space of horospheres labelled by 
$(\xi,\lambda)$,

$$\eqalign{F^a_{r\mu}(g\cdot \xi, g\cdot \lambda)~&=~
\int_{H^4}~dx~\langle 0 \vert~[j_\mu^a(x), \phi_r]~\vert 0 \rangle 
~[x, g\cdot \xi]^{-i(g\cdot \lambda)-{3\over 2}}
\cr
~&=~\int_{H^4}~dx~\langle 0 \vert ~[j_\mu^a(x),\phi_r]~\vert 0 \rangle
~[g^{-1}\cdot x, \xi]^{-i(g\cdot \lambda)-{3\over 2}}
\cr
~&=~\int_{H^4}~dx~\langle 0 \vert ~[j_\mu^a(g\cdot x), \phi_r]~\vert 0 \rangle
~[x,\xi]^{-i(g\cdot \lambda)-{3\over 2}}
\cr
~&=~(g)_\mu^\nu~\int_{H^4}~dx~\langle 0\vert ~[j_\nu^a(x), \phi_r]~\vert 0 
\rangle~[x,\xi]^{-i(g\cdot \lambda)-{3\over 2}}
\cr}
\eqno(36)
$$

Since the elements $g\in SO(4)$ translate points on $S^3$ and leave $\lambda$
invariant,

$$F^a_{r\mu}(g\cdot \xi, g\cdot \lambda)~=~(g)_\mu^\nu~
\int_{H^4}~dx~\langle 0\vert~[j_\nu^a(x), \phi_r]~\vert 0 \rangle 
~[x, \xi]^{-i\lambda-{3\over 2}}
~=~(g)_\mu^\nu~F^a_{r\nu}(\xi,\lambda)
\eqno(37)$$

Since the expectation value of the commutator transforms covariantly under the
action of the subgroup SO(4) of the isometry group $H^4$, it follows that

$$2~F_{[\rho;\lambda]}(g \cdot y, g\cdot y^\prime)~=
~2(g)_\rho^\nu (g)_\lambda^\sigma~F_{[\nu;\sigma]}(y,y^\prime)=~2~(g)_\rho^\nu~
F_{[\nu;\lambda]}(y, y^\prime)
\eqno(38)$$

Consequently, $F_{[\mu;\sigma]}(y,y^\prime)~=~0$ and 
$F_\mu(y,y^\prime)~=~\partial_\mu D(y,y^\prime)$.  Although 
$F^a_{r\mu}~=~\partial_\mu D(x)$ does not imply $F^a_{r\mu}~=~-iq_\mu D(q)$,
integration by parts can be used to show that

$$\eqalign{\int_{H^4}~d^4x~{\sqrt{g(x)}}~\nabla^\mu \partial_\mu D(x)~
[x,\xi]^{-i\lambda-{3\over 2}}
~&=~\int_{H^4}~d^4x~{\sqrt {g(x)}}~D(x)~\nabla^\mu \partial_\mu~
[x, \xi]^{-i\lambda-{3\over 2}}
\cr
~&=~-(\lambda^2~+~{9\over 4})~\int_{H^4}~dx~D(x)
~[x, \xi]^{-i\lambda-{3\over 2}}
\cr
~&=~-(\lambda^2~+~{9\over 4})~D(\xi,\lambda)
\cr}
\eqno(39)
$$

Replacing $\lambda$ by $-i\lambda$ for the analytic continuation to 
anti-de Sitter space and using current conservation implies that
$D(\xi,\lambda)$ contains $\delta(\lambda^2-{9\over 4})$.  The physical
spectrum is given by a discrete series representation of SO(3,2) with 
the eigenvalues of the Casimir operator equal to $-\omega_0(\omega_0-3)$,
so that $\omega_0~=~\lambda+{3\over 2}$.  Thus the condition 
$\lambda~=~{3\over 2}$ is equivalent to $\omega_0~=~3$.  As representations
of SO(3,2) are denoted as ${\cal D}(\omega_0,s)$ 
where $\omega_0~\ge~s+{1\over 2}$ is the lowest energy eigenvalue and 
$s$ is the spin of the field, the
Goldstone bosons correspond to the representation ${\cal D}(3,0)$. 

\noindent
{\bf 8. Equivalence of Vacua in Different Coordinate Systems}

A closer analogy with the construction of Feynman diagrams in Minkowski 
space-time can be achieved by using conformally flat coordinates in anti-de 
Sitter space, as there are flat three-dimensional sections of the space-time 
that are spanned by three of the coordinates.  Plane waves in three 
dimensions $e^{i(k_{t^\prime}t^\prime-k_yy-k_zz)}$ are then included in the 
eigenfunctions leading to
an identification of the transform variables with components of the 
momentum.  Conservation of momentum follows from the delta functions that
arise in the conversion of the configuration space integrals to
the transform space integrals.

To establish  the connection between the analytic continuation of the 
Feynman propagator (11) and the propagator $\langle 0\vert T \phi(x)
\phi(x^\prime) \vert 0\rangle$ in these coordinates, it is sufficient
to establish equivalence of the vacua in the different coordinate systems.
From the embedding of anti-de Sitter space as a hyperboloid in the 
five-dimensional pseudo-Euclidean space with signature $(+---+)$, one obtains
the metric in different intrinsic coordinates and the corresponding 
eigenfunctions of the d'Alembertian.

A brief list of coordinate systems and eigenfunctions associated with a 
special choice of eigenvalue is now given.
\vskip 5pt
\noindent
(i)  Globally static coordinates 
$$\eqalign{z_0~&=~(a^2+r^2)^{1\over 2}~ sin ({t\over a})
\cr
z_1~&=~r~ sin~\theta~ cos~ \phi
\cr
z_2~&=~r~ sin ~\theta~ sin~ \phi
\cr
z_3~&=~r~ cos~ \theta
\cr
z_4~&=~(a^2+r^2)^{1\over 2}~cos({t\over a})
\cr}
\eqno(40)$$
$$ds^2~=~\left(1+{{r^2}\over {a^2}}\right)dt^2-\left(1+{{r^2}\over {a^2}}
\right)^{-1}
dr^2-r^2(d\theta^2+sin^2\theta d\phi^2)
\eqno(41)$$
Since the curvature scalar is $-{{12}\over {a^2}}$, the conformally invariant
scalar field satisfies
$$\eqalign{\left(1+{{r^2}\over {a^2}}\right)^{-1}{{\partial^2 \Phi}\over 
{\partial t^2}}
-{1\over {r^2}}{\partial\over {\partial r}} \left[r^2\left(1+{{r^2}\over {a^2}}
\right)
{{\partial\Phi}\over {\partial r}}\right]-{1\over {r^2sin~\theta}}
{\partial\over
{\partial\theta}}&\left[sin~\theta{{\partial\Phi}\over {\partial\theta}}\right]
\cr
&-{1\over {r^2 sin~\theta}}{{\partial^2 \Phi}\over {\partial\phi^2}}-
{2\over {a^2}}\Phi~=~0
\cr}
\eqno(42)$$
The basis set of solutions are
$$\eqalign{\phi_{\omega l m}~&=~{{f_l(r)}\over r}Y_{l m}(\theta, \phi)
e^{-i\omega t}
\cr
f_l^{(1)}(r)~&=~e^{-i\pi a \omega}2^{-l-1} \pi^{-{1\over 2}}\Gamma(l+1-a\omega)
({{ir}\over a})^{l+1} (1+{{r^2}\over {a^2}})^{-{{a\omega}\over 2}}
\cr
&~~~~~~~F(1+{l\over 2}-{{a\omega}\over 2},{1\over 2}+{l\over 2}-{{a\omega}
\over 2};l+{3\over 2};-{{r^2}\over {a^2}})
\cr
f_l^{(2)}(r)~&=~{{\Gamma(-{1\over 2}-l)({{ir}\over a})^{l+1}}\over 
{2^{l+1}\pi^{1\over 2}(1+{{r^2}\over {a^2}})^{-{{a\omega}\over 2}}
\Gamma(-l+a\omega)}} 
F({1\over 2}+{l\over 2}+{{a\omega}\over 2}, 1+{l\over 2}
+{{a\omega}\over 2};l+{3\over 2};-{{r^2}\over {a^2}})
\cr
&~~~~~~~~~+~{{2^l \pi^{-{1\over 2}}\Gamma({1\over 2}+l)({{ir}\over a})^{-l}}
\over {(1+{{r^2}\over {a^2}})^{-{{a\omega}\over 2}}\Gamma(1+l+a\omega)}}
F(-{l\over 2}+{{a\omega}\over 2}, {1\over 2}-{l\over 2}+{{a\omega}\over 2};
{1\over 2}-l;-{{r^2}\over {a^2}})
\cr}
\eqno(43)$$
The general solution will involve the combination $A_lf_l^{(1)}+B_lf_l^{(2)}$ 
and the choice of $A_l$ and $B_l$ correspond to the choice of vacuum state.
Defining the scalar product in the space of solutions to be
$$(\phi_1,\phi_2)~=~i\int d^3 x {\sqrt{-g}} g^{0\nu} ({\bar\phi_1}
\partial_\nu \phi_2~-~\phi_2 \partial_\nu {\bar \phi_1})
\eqno(44)$$
finiteness of the norm at $r$=0 implies that $B_l=0$ because the second term in
$f_l^{(2)}$ contains a factor of $r^{-l}$.

These coordinates can be obtained by analytic continuation of globally static 
coordinates in de Sitter space after mapping $a\to -ia$.  The de Sitter 
solutions are
$$\eqalign{\phi_{\omega l m}^{dS}~&=~{{g_l(r)}\over r} Y_{lm}(\theta,\phi) 
e^{-i\omega t}
\cr
g_l^{(1)}(r)~&=~Q_l^{ia\omega}({a\over r})~~~~~g_l^{(2)}(r)~=~
{\cal P}_l^{ia\omega}({a\over r})
\cr}
\eqno(45)$$
Since $(\phi_{\omega l m}^{dS}, \phi_{\omega l m}^{dS})$ is finite at $r$=0 
only if $Q_l^{ia\omega}({a\over r})$ is chosen for $g_l(r)$, analytic 
continuation to AdS gives the same mode solutions.  

A similar set of coordinates exist on $S^4$ and analytic continuation to AdS
can be achieved through the mapping of the radial coordinate $r\to ir$.  
Again, finiteness of $(\phi_{\omega l m}^{S^4}, \phi_{\omega l m}^{S^4})$
leads to the same solutions after analytic continuation.

The propagator for a massive scalar field can also be continued from
$S^4$ to anti-de Sitter space, and it has been noted that this is
not equal to the direct mode sum in anti-de Sitter space, where the modes
are subject to supersymmetric boundary conditions [11] .  
The latter propagator is the sum of a symmetric and an anti-symmetric 
combination of the analytically continued $S^4$ propagator 
and its antipodal counterpart.   

It may be recalled that the basis solutions $\phi_{\omega l m}$ can be split
into two groups, corresponding to Dirichlet and Neumann boundary conditions.
The conformally coupled scalar field modes form the representations
${\cal D}(1,0)$ and ${\cal D}(2,0)$ respectively.  Instead of using 
${\cal D}(1,0)$ or ${\cal D}(2,0)$, the two representations can be combined
to give two irreducible representations with different parity assignments [12].
$$\eqalign{{\cal D}(1,0)^+:~~\omega~&=~n+l+1,~~n~=~0,~1,~2,...,~~
l~=~0,~1,~2,~...~~~~~~parity~=~(-1)^{\omega-1}
\cr
A_{\omega lm}~&=~\phi_{\omega lm}~~~~~~~B_{\omega lm}~=~0~~~~~~~~~~~~~~~~even~n
\cr
A_{\omega lm}~&=~0~~~~~~~~~~~~B_{\omega lm}~=~i \phi_{\omega lm}~~~~~~~~~~~
odd~n
\cr
{\cal D}(1,0)^-:~~\omega~&=~n+l+1~~~n~=~0,~1,~2,...~~~l~=~0,~1,~2,...
~~~~~~~parity~=~-(-1)^{\omega-1}
\cr
A_{\omega lm}~&=~0~~~~~~~~~~~~~B_{\omega lm}~=~\phi_{\omega lm}~~~~~~~~~~even~n
\cr
A_{\omega lm}~&=~-i\phi_{\omega lm}~~~~~B_{\omega lm}~=~0~~~~~~~~~~~~~~~~odd~n
\cr}
\eqno(46)
$$
where $A$ and $B$ are scalar and pseudo-scalar fields respectively.
A study of the spin-${1\over 2}$ field reveals that the imposition of
Dirichlet and Neumann conditions again leads to quantization of the
frequencies.  The basis solutions are
$\{\chi_{\omega jm}^{(+)}\},
~\omega~=~2n+j+1,~j~=~{1\over 2},~{3\over 2},~...$ 
with parity $(-1)^{j-{1\over 2}}$, and 
$\{\chi_{\omega jm}^{(-)}
~=~i\gamma_5 \chi_{\omega jm}^{(+)}\},~\omega~=~2n+j+1,
~j~=~{1\over 2},~{3\over 2},~...$ with parity $-(-1)^{j-{1\over 2}}$.
Again, these two representations can be combined into two irreducible
representations with different parity assignments.
$$\eqalign{{\cal D}({3\over 2},{1\over 2})^+:~\omega~&=~n+j+1,~~
n~=~0,~1,~2,~...,~j~=~{1\over 2},~{3\over 2},~{5\over 2},~...
~~~parity=(-1)^{\omega-{3\over 2}}
\cr
\chi_{\omega jm}^+~&=~i\chi_{\omega jm}^{(+)}~~~~~~~~~~~~~even~n
\cr
&~~~~~\chi_{\omega jm}^{(-)}~~~~~~~~~~~~~~~odd~n
\cr
{\cal D}({3\over 2}, {1\over 2})^-:~\omega~&=~n+j+1,~n~=~0,~1,~2,~...,
~j~=~{1\over 2},~{3\over 2},~{5\over 2},~...~
~~parity=-(-1)^{\omega-{3\over 2}}
\cr
\chi_{\omega jm}^-~&=~i\chi_{\omega jm}^{(-)}~~~~~~~~~~~~even~n
\cr
&~~~-\chi_{\omega jm}^{(+)}~~~~~~~~~~~~~odd~n
\cr}
\eqno(47)
$$
The representations ${\cal D}(1,0)^+$ and ${\cal D}(1,0)^-$ can be related
to the representations ${\cal D}({3\over 2},{1\over 2})^+$ and 
${\cal D}({3\over 2},{1\over 2})^-$ by the supersymmetry transformation
$$\eqalign{\delta A~&=~{1\over {\sqrt 2}}~{\bar \epsilon}\chi
~~~~~~~\delta B~=~{i\over {\sqrt 2}}{\bar \epsilon}\gamma_5 \chi
\cr
\delta\chi~&=~-{1\over {\sqrt 2}}~[i \gamma^\mu\partial_\mu(A+i\gamma_5B)
~+~a(A-i\gamma_5B)]
\cr}
\eqno(48)
$$
For a general spin $s$, two unitary irreducible representations 
${\cal D}(s+1,s)^{\pm},~\omega~=~n+l+s+1$ can be formed, [12][13] and the
sets 
$\{{\cal D}(1,0)^+, {\cal D}({3\over 2}, {1\over 2})^+,{\cal D}(2,1)^+,~...\}$
and 
$\{{\cal D}(1,0)^-, {\cal D}({3\over 2}, {1\over 2})^-, {\cal D}(2,1)^-,~...\}$
are separately invariant under supersymmetry transformations.

It has been indicated already that analytic continuation from $H^4$ or $S^4$
to anti-de Sitter space could be used to obtain an expression for the 
Green function.  The lack of global hyperbolicity of anti-de Sitter space 
requires the use of reflective boundary conditions, which are included in the 
supersymmetric boundary conditions above, and their effect
can be reproduced on $H^4$ by adding a second source on the lower sheet of
the hyperboloid.  The analyticity properties of the Green function in
de Sitter space [14] or Euclidean de Sitter space, $S^4$, lead to
constraints on the parity, which explains the discrepancy between the 
anti-de Sitter propagator associated with supersymmetric boundary 
conditions and the propagator continued from $S^4$.     

{\noindent{(ii)}} Conformal mapping from the Einstein static universe
$$\eqalign{z_0~&=~a~sin~t~sec~\rho
\cr
z_1~&=~a~tan~\rho~sin~\theta~cos~\phi
\cr
z_2~&=~a~tan~\rho~sin~\theta~sin~\phi    
\cr
z_3~&=~a~tan~\rho~cos~\theta
\cr
z_4~&=~a~cos~t~sec~\rho
\cr}
\eqno(49)$$
$$ds^2~=~{{a^2}\over {cos^2\rho}}~(dt^2-d\rho^2-sin^2\rho
~(d\theta^2+sin^2 d\phi^2))~=~{{a^2}\over {cos^2\rho}}ds_E^2
\eqno(50)$$
where $ds_E^2$ is the line element for the Einstein static universe $R^1\times
S^3$.  The solution to the conformally invariant scalar field equation can be
can be found by a conformal mapping of the solution in the Einstein static 
universe [15].
$$\eqalign{\phi_{\omega l m}^{AdS}~&=(cos \rho)\phi_{\omega l m}^E=
N_{\omega l}e^{-i\omega t}(sin\rho)^l cos\rho
\cr
&~~~~~~~~~~~~~~~~~~~~~~~~~ F({1\over 2}(l+1-\omega), 
{1\over 2}(l+1+\omega);l+{3\over 2};sin^2 \rho) Y_{lm}(\theta,\phi)
\cr
\omega~&=~2j+l+1,~2j+l+2,~j~=~0,1,2,...
\cr}
\eqno(51)$$
where $N_{\omega l}$ is a normalization factor.  A second set of 
solutions to  the conformally invariant wave equation would contain 
$(sin^2\rho)^{-l-{1\over 2}}F({1\over 2}(\omega -l), 
-{1\over 2}(\omega+l);{1\over 2};cos^2\rho)$.
Finiteness of $(\phi_{\omega l m}, \phi_{\omega l m})$ at $\rho=0$ excludes
this solution. 

The first set of solutions in equation (43) and the functions in equation (51)
\hfil\break
can be shown to be equivalent using the transformation 
$r\to a~tan~\rho,~{t\over a}\to t$ from 
\hfil\break
globally static coordinates to those of the Einstein static universe and 
the identity
\hfil\break
$F(a,b;c;x)~=~(1-x)^{-b} F(b, c-a;c;{x\over {x-1}})$.  Solutions 
to the general Klein-Gordon equation will have the same form as the solutions
in equation (51), except that $\omega=\omega_0+2j+l$ where $\omega_0$ depends
on the mass.
 
\vfill\eject
{\noindent{(iii)}} Conformally flat coordinates

$$\eqalign{z_0-z_1~&=~\rho^\prime+{1\over \rho^\prime}(y^2+z^2-t^{\prime 2})
\cr
z_0+z_1~&=~{1\over {a^2\rho^\prime}}
\cr
z_2~&=~{y\over {a\rho^\prime}} 
\cr
z_3-z_4~&=~{{z+t^\prime}\over {a\rho^\prime}}
\cr
z_3+z_4~&=~{{z-t^\prime}\over {a\rho^\prime}}
\cr}
\eqno(52)$$

$$ds^2~=~{1\over {a^2\rho^{\prime 2}}}~[dt^{\prime 2}-d\rho^{\prime 2}-
dy^2-dz^2]
\eqno(53)$$
covers half of the space, with $\rho > 0$.  The other half is covered by
choosing $\rho < 0$.
The Klein-Gordon equation 
$$-a^2\rho^{\prime 2}{{\partial^2\Phi}\over {\partial\rho^{\prime 2}}}
+2a^2\rho^\prime {{\partial\Phi}\over {\partial\rho^\prime}}
-a^2\rho^{\prime 2}\left({{\partial^2\phi}\over {\partial y^2}}+
{{\partial^2\Phi}\over {\partial z^2}}-{{\partial^2\Phi}
\over {\partial t^{\prime 2}}}\right) +m^2\Phi~=~0
\eqno(54)$$
has basis solutions
$$\eqalign{
\phi_{k_{t^\prime},k_y,k_z}(t^\prime,\rho^\prime,y,z)~=~
{a\over {2{\sqrt 2}\pi}}~&
{\sqrt {1~-~{{(k_y^2+k_z^2)}\over {k_{t^\prime}^2}}} } 
~e^{-i(k_{t^\prime}t^\prime-k_yy-k_zz)}
\cr
&\rho^{\prime{3\over 2}}~J_{\sqrt{{{m^2}\over {a^2}}+{9\over 4}}}
({\sqrt{{k_{t^\prime}}^2-{k_y}^2-{k_z}^2}}\rho^\prime)
\cr}
\eqno(55)$$
Note that $\phi_{k_{t^\prime},k_y,k_z}(t^\prime, \rho^\prime, y,z)$ 
contains a factor representing a plane wave in
the three dimensional subspaces  spanned by the coordinates $(t^\prime,y,z)$.

The mode solutions are associated with a choice of vacuum state for the
scalar field, and it is of interest to determine whether the vacuum states are
equivalent in the three coordinate systems for anti-de Sitter space.  A
mixing of positive and negative frequencies in the transformation between
coordinate systems would change a vacuum state into one with non-zero particle
number.

Since the time coordinate differs by a constant scaling factor for coordinate
systems (i) and (ii), it follows trivially that the vacuum states in these
coordinates are the same.  To compare the vacuum states in coordinate systems
(ii) and (iii), it is useful to expand the solutions 
in equation (55) in terms of the solutions
in equation (51), with $\omega=\omega_0+2j+l$.
$$\eqalign{e^{-ik_{t^\prime}t^\prime}e^{ik_yy}e^{ik_z z}
\rho^{\prime{3\over 2}} J_\nu(-ik\rho^\prime)
\cr
~&=~\sum_{\omega l m} \alpha_{{{\omega lm}\atop{k_{t^\prime}k_yk_z}}}
e^{-i\omega t}Y_{lm}(\theta\phi)(sin~\rho)^l cos~\rho f_{\omega l}(\rho)
\cr
&~~~+~\sum_{\omega lm} \beta_{{{\omega lm}\atop {k_{t^\prime}k_yk_z}}}
e^{i\omega t}Y_{lm}^\ast(\theta,\phi)(sin~\rho)^lcos~\rho f_{\omega l}(\rho)
\cr}
\eqno(56)$$
Since
$$\eqalign{{1\over {2\pi}}\int_{-\pi}^\pi& e^{i\omega^\prime t}
e^{-ik_{t^\prime}t^\prime}
e^{ik_yy}e^{ik_zz}\rho^{\prime{3\over 2}} J_\nu(-ik\rho^\prime)dt
\cr
&=\sum_{lm} \alpha_{{{\omega^\prime lm}\atop {k_{t^\prime}k_yk_z}}}
Y_{lm}(\theta\phi)
(sin~\rho)^l cos~\rho f_{\omega^\prime l}(\rho) \theta(\omega^\prime)
\cr
&~~~~+\sum_{lm} \beta_{{{-\omega^\prime lm}\atop {k_{t^\prime}k_yk_z}}}
Y_{lm}^\ast(\theta,\phi) (sin~\rho)^l cos~\rho f_{-\omega^\prime l}(\rho) 
\theta(-\omega^\prime)
\cr}
\eqno(57)$$
vanishing of the integral implies that 
$\beta_{{{-\omega^\prime lm}\atop {k_{t^\prime}k_yk_z}}}=0$ for all 
$\omega^\prime, l, m$ because the 
functions
\hfil
\break
$Y_{lm}^\ast(\theta\phi) (sin~\rho)^l cos~\rho f_{-\omega^\prime l}(\rho)$ form
a complete basis in three dimensions.  This method therefore requires
evaluation of the integral
$${1\over {2\pi}}\int_{-\pi}^\pi~dt~e^{i\omega^\prime 
arc~tan~[{a\over {2t^\prime}}
[{1\over {a^2}}+\rho^{\prime 2}+y^2+z^2-t^{\prime 2}]]}
e^{-ik_{t^\prime}t^\prime}e^{ik_yy}e^{ik_zz} \rho^{\prime{3\over 2}}
J_\nu(-ik\rho^\prime)
\eqno(58)$$

Equivalence of the vacua in coordinates of the Einstein static universe
$(t,\rho, \theta,\phi)$ and the conformally flat coordinates 
$(t^\prime,\rho^\prime,y,z)$ can be demonstrated more easily by showing that 
the domain of holomorphicity of $e^{-i\omega t}$ is mapped into the domain of
holomorphicity of $e^{-i\omega t^\prime}$.
Let $t=\tau+i\sigma,~-\pi\le\tau<\pi,~0\le\sigma<\infty$.  The time coordinate
$t^\prime$ is given by
$$t^\prime~=~-{{cos(\tau+i\sigma)sec~\rho}\over
{a[sin(\tau+i\sigma)sec~\rho+tan~\rho~sin~\theta~cos~\phi]}}
\eqno(59)$$
When
$$\eqalign{I.~~ \sigma~&=~0~~~~(-\pi,\pi)\to(-\infty,\infty)
\cr
II.~~ \sigma~&>~0
\cr
Im~&t^\prime~=~{1\over a}{{sinh~\sigma(cosh~\sigma ~sec^2\rho+sin~\tau~sec~\rho
~tan~\rho~sin~\theta~cos~\phi)}
\over {(sin~\tau~cosh~\sigma~sec~\rho+tan~\rho~sin~\theta~cos~\phi)^2
+cos^2\tau~sinh^2\sigma~sec^2\rho}}~ >~0
\cr
as&~0\le\rho\le {\pi\over 2},~ 0\le\theta\le\pi,~0\le\phi\le 2\pi
\cr
III.~~\sigma~&<0~~~~Im~t^\prime~<~0
\cr}
\eqno(60)$$
From equation (60), it follows that the domains of holomorphicity 
$e^{-i\omega t}$ and $e^{-i\omega t^\prime}$ are mapped into each other and 
there is
no mixing of positive and negative frequencies in the conformally flat 
coordinates.  Thus, the vacuum state for the scalar field in all three 
coordinate systems will be equivalent.

Configuration and momentum space Feynman rules can again be formulated in 
these coordinates, and they closely resemble the flat-space Feynman rules
as the eigenfunctions of the Laplacian contain three-dimensional
plane waves, which give rise to delta functions representing conservation
of momentum in the three dimensions.  
The factor $\prod_{vert.}~f^{\{k\}}(vert.)$ is now determined by the integral
of the product of Bessel functions.
Identities such as
$$\eqalign{\int_0^\infty~d\rho~\rho^{r-1}~
J_\nu(a\rho)~J_\nu(b\rho)~J_\nu(c\rho)
~&=~{{2^{r-1}~(ab)^\nu~c^{-2\nu-r}~\Gamma\left({{3\nu+r}\over 2}\right)}
\over {(\Gamma(\nu+1))^2~\Gamma\left(1-{{\nu+r}\over 2}\right)}}
\cr
&~\cdot~~F_4\left({{\nu+r}\over 2}, {{3\nu+r}\over 2};\nu+1,\nu+1;
{{a^2}\over {c^2}};{{b^2}\over {c^2}}\right)
\cr
Re~(3\nu+r)~>~0~~~~Re~r~<~{5\over 2}&~~~~a~>~0,~~b~>~0,~~c~>~0,~~
c~>~a+b
\cr
\int_0^\infty~d\rho~\rho^{r-1}~J_r(cx)~\prod_{i=1}^n~\rho_i^{-\mu_i}
~J_{\mu_i}(a_i \rho_i)
~&=~2^{r-1}~\Gamma(r)~c^{-r}~\prod_{i=1}^n~[~b_i^{-\mu_i}~J_{\mu_i}(a_i b_i)]
\cr
\rho_i~=~{\sqrt{\rho^2+b_i^2}}~~~a_i~=~0~~~Re~b_i~>0~~~&\sum_{i=1}^n~a_i~<~c
\cr
Re~\left({1\over 2}n~+~\sum_{i=1}^n \mu_i~+~{3\over 2}\right)~
&>~Re~r~>~0
\cr}
\eqno(61)
$$
reveal that the Bessel function integrals can be evaluated, although they
no longer contain the delta functions associated with the
orthogonality relations for two Bessel functions [16].  Thus, the vertex
factor $\prod_{vert.}~f^{\{k\}}(vert.)$ will give rise to extra non-trivial
functions of the momenta which must be included in the momentum-space
integrals.

\vfill\eject
\noindent
{\bf 9. Lehmann Representation for the Propagator in the Presence} 
\hfil\break
{\bf {\phantom {....}} of a Second Source in $H^4$}

For an interacting scalar field, the two-point function can be
expressed in terms of the free-field Green functions
$$\eqalign{\langle 0 \vert T \phi(x) \phi(x^\prime) \vert 0 \rangle
~&=~\sum_{\omega=\omega_{0\pm}}^\infty~\rho(\omega, \phi) 
G_\omega(x,x^\prime)
\cr
&~~~~~\omega~=~\omega_{0\pm}, \omega_{0\pm}+1 , ...
\cr}
\eqno(62)
$$
where $\omega_{0\pm}$ are the lowest eigenvalues of $J_{04}$, the energy
operator, and
$$\eqalign{J_{AB}~&=~\int~d^3 x~{\sqrt{-g}}~T^0_\nu K^\nu_{AB} 
\cr
K_{AB}~&=~y_A~{{\partial}\over {\partial y^B}}~-
             ~y_B~{{\partial}\over {\partial y^A}}
\cr
\eta_{AB}y^A y^B~&=~{1\over {a^2}}
\cr}
\eqno(63)
$$
The Green function $G_{\omega \pm}(x,x^\prime)$ satisfies the equation
$$\eqalign{(\nabla^\mu \nabla_\mu~+~m^2)G_{\omega_{0\pm}}(x,x^\prime)
~&=~{1\over {\sqrt{-g}}}~\delta(x,x^\prime)
\cr
\omega_{0\pm}~&=~{3\over 2}~\pm~{\sqrt{{9 \over 4}~+~{{m^2}\over {a^2}}}}
\cr}
\eqno(64)
$$
The curved-space delta function 
${\tilde \delta}(x,x^\prime)~=~{1\over {\sqrt{-g}}}~\delta(x,x^\prime)$ 
is defined so that
\hfil\break
$\int_X~d^4x~{\sqrt {-g}}~{\tilde \delta(x,x^\prime)}=1~=~
\int~d^4x {\sqrt {-g}}~{1\over {\sqrt {-g}}}~\delta(x,x^\prime)$.
The Lehmann spectral representation for the two-point function can be 
deduced from the Fourier transformation of the analytically continued
Green function in $H^4$. Reflective boundary conditions may be realized
in $H^4$ by adding a second source for the Green function at the anti-podal
point [$x^A \to -x^A$] on the second sheet of the hyperboloid.
$$(\nabla^\mu \nabla_\mu~+~m^2)~G_{\omega_{0\pm}}^{E.R.B.}(x,x^\prime)~=~
{1\over {\sqrt g}}~[\delta(x,x^\prime)\pm \delta(x,{\hat x}^\prime)]
\eqno(65)
$$
The solution to this equation is
$$G_{\omega_{0\pm}}^{E.R.B.}(x,x^\prime)~=~G_{\omega_{0\pm}}^E(x,x^\prime) ~\pm
~G_{\omega_{0\pm}}^E(x,{\hat x}^\prime) 
\eqno(66)
$$
For a conformally coupled massless scalar field with $m^2~=-2a^2$, 
$\omega_{0+}~=~2$ and $\omega_{0-}~=~1$, the Green function satisfying
reflective boundary conditions is
$$G_{\omega_{0\pm}}^{E.R.B.}(u_E)~=~{1 \over {8 \pi^2}}
~\left[{1\over {u_E}}~\pm {1\over {2-u_E}}~\right]
\eqno(67)
$$
and
$$G_{\omega_{0+}}^{E.R.B.}(x,x^\prime)~\approx~-{1\over {4\pi^2}}
{1\over {u_E^2}}
~~~~~G_{\omega_{0-}}^{E.R.B.}(x,x^\prime)~\approx~{1\over {4 \pi^2}}
{1\over {u_E}}
\eqno(68)
$$
as $u_E~=~{1\over 2} a^2 (x^A-x^{\prime A})^2~\to~\infty$.    
Defining $z$ to be $x^A x^\prime_A$, the Fourier representation [4] is
$$G_\omega^E(z)~=~{i\over 2}~
\int_{-{1\over 2}+i\infty}^{{1\over 2}+i\infty}~d\sigma~(\sigma+1)\sigma
(\sigma-1)~ctg~\pi\sigma~{\tilde G}^E_\omega(\sigma) 
{{{\cal P}_\sigma^{-1}(z)}
\over {(z^2-1)^{1\over 2}}}
\eqno(69)
$$
where
$${\tilde G}^E_\omega(\sigma)~=~\int_1^\infty~dz~G^E_\lambda(z)
(z^2~-~1)^{1\over 2} {\cal P}^{-1}_\sigma(z)
\eqno(70)
$$
and ${{{\cal P}^{-1}_\sigma(z)}\over {(z^2-1)^{1\over 2}}}$ being the 
eigenfunction of $\nabla^\mu \nabla_\mu$ with eigenvalue 
$-(\sigma(\sigma+1)~-~2)$.
The integral (70) represents integration over the upper sheet of the
hyperboloid.  The other possible range of the argument of the Green function
corresponds to integration over the second sheet, with $u_E~\ge~2$, 
and ${\hat z}~=~x^A~{\hat x}^\prime_A$ ranging from $-1$ to $-\infty$.
It follows that
$$\eqalign{G_{\omega_{0\pm}}^{E.R.B.}(z)~&=~G_{\omega_{0\pm}}^E(x,x^\prime)\pm
G_{\omega_{0\pm}}^E(x,{\hat x}^\prime)
\cr
&={i\over 2}~\int_{-{1\over 2}-{i\infty}}^{-{1\over 2}+i\infty} 
d\sigma~(\sigma+1)\sigma(\sigma-1)~ctg~\pi\sigma 
{\tilde G}^E_{\omega_{0\pm}}(\sigma)
{{{\cal P}^{-1}_\sigma(z)}\over {(z^2-1)^{1\over 2}}}
\cr
&~\pm~{i\over 2}~\int_{-{1\over 2}~-~i\infty}^{-{1\over 2}~+~i\infty}
d \sigma~(\sigma+1)\sigma (\sigma-1)~ctg~\pi\sigma~
{\hat G}_{\omega_{0\pm}}^E(\sigma){{{\cal P}_\sigma^{-1}({\hat z})}\over 
{({\hat z}^2-1)^{1\over 2}}}
\cr}
\eqno(71)
$$
where
$${\hat G}_{\lambda\pm}^E(\sigma)~=~\int_{-1}^{-\infty}~d{\hat z}~
G_{\omega_{0\pm}}^E({\hat z})({\hat z}^2-1)^{1\over 2}~
{\cal P}^{-1}_\sigma({\hat z})
\eqno(72)
$$ 
Applying the Klein-Gordon operator to the Green function (71) and requiring
that it satisfy the equation (65) implies that 
$${\hat G}^E_{\omega_{0\pm}}(\sigma)~=~{1\over {\omega_{0\pm}(\omega_{0\pm}-3)
-\sigma(\sigma+1)+2}}
\eqno(73)
$$
and thus the propagator in transform space is essentially unchanged from
the one given in equation (23).

\vfill\eject
\noindent
{\bf 10. Curvature, Shifts in the Momentum and Ground State}
\hfil\break 
{\bf {\phantom {......}} Contributions to the String Hamiltonian}

The definition of momentum might also be considered within the context
of the effect of curvature on the commutation relations for position and
momentum operators [17].  If the momentum operators are defined so that 
they induce the change in geodesic coordinates from  
$x_Q^\mu$ to $x_{Q^\prime}^\mu$, then
$$x_Q^\mu~\to~x_{Q^\prime}^\mu~=~x_Q^\mu~-~{1\over {i{\hbar}}}\alpha^\nu
[x_Q^\mu, p_{Q\nu}]
\eqno(74)
$$
Assuming that a local coordinate system has been chosen so that
$\Gamma^\mu_{\alpha\beta}(Q)~=~0$, derivatives of the geodesic equation
imply that
$$\eqalign{x_{Q^\prime}^\mu~&=~x_Q^\mu~-~\alpha^\mu~+~{1\over 2} 
\Gamma^\mu_{\alpha\beta,\nu}(Q)~\alpha^\nu x_Q^\alpha x_Q^\beta
~+~{\cal O}(\alpha^2,x_Q^3)
\cr
~&=~x_Q^\mu~-~\alpha^\mu~-~{1\over 6}(R^\mu{}_{\alpha\beta\nu}(Q)~+~
R^\mu{}_{\beta\alpha\nu}(Q))~\alpha^\nu x_Q^\alpha x_Q^\beta
\cr}
\eqno(75)
$$    
The momentum operator which generates such a translation of the coordinate
$x^\rho$ is
$$-i {\partial\over {\partial x^\nu}}~-~
{i\over 6}\left(R^\rho{}_{\alpha\beta\nu}~+~R^\rho{}_{\beta\alpha\nu}\right)
x^\alpha x^\beta {\partial\over {\partial x^\rho}}
\eqno(76)
$$
which gives
$$k_\nu~+~{1\over 6}\left[R^\rho{}_{\alpha\beta\nu}
~+~R^\rho{}_{\beta\alpha\nu} \right]~x^\alpha x^\beta k_\rho
\eqno(77)
$$
after applying the operator to the plane wave $e^{ik\cdot x}$, whose
use would be justified by the approximate flatness of the space-time
outside the local region with curvature.  The result
depends on the choice of coordinate $x^\rho$ and the index $\rho$ is not
summed even though it occurs more than once in the formula (77).  To
obtain a directionally-averaged definition of the momentum, one may
sum over $\rho$ and divide by four to obtain the following result 
$$k^\prime_\mu~=~k_\mu~+~{1\over {24}}~\sum_\rho~[R^\rho{}_{\alpha\beta\mu}
~+~R^\rho{}_{\beta\alpha\mu}]~x^\alpha x^\beta k_\rho
\eqno(78)
$$
For a constant curvature metric, such as the one that is used for 
anti-de Sitter space, the identity
$$R_{\rho\alpha\beta\nu}~+~R_{\rho\beta\alpha\nu}~=~
-a^2(g_{\rho\beta}~g_{\alpha\nu}~+~g_{\rho\alpha}g_{\beta\nu}
~-~2g_{\rho\nu}g_{\alpha\beta})
\eqno(79)
$$
is valid.  
$$\eqalign{k^{\prime 2}~&=~k^2~-~{{a^2}\over {12}}~(g_{\rho\beta}g_{\alpha\nu}
~+~g_{\rho\alpha}g_{\beta\nu}~-~2g_{\rho\nu}g_{\alpha\beta})k^\rho k^\nu
x^\alpha x^\beta
\cr
~&+~{{a^4}\over {576}}(g_{\rho\beta}g_{\alpha\nu}~+~g_{\rho\alpha}g_{\beta\nu}
~-~2g_{\rho\nu} g_{\alpha\beta})(g_{\sigma\delta}\delta_\gamma^\nu~+
~g_{\sigma\gamma}\delta_\delta^\nu~-~2\delta_\sigma^\nu g_{\gamma\delta})
x^\alpha x^\beta x^\gamma x^\delta k^\rho k^\sigma
\cr
\langle k^{\prime 2} \rangle~&=~k^2 ~\left[~1~+~{{a^2}\over 8}
\langle x^2\rangle~+~ {{a^4}\over {192}}~\langle x^4 \rangle~\right]
\cr}
\eqno(80)
$$

If the index $\rho$ in equation (76) is summed, then the shift in (75) is
not altered, but each directional derivative would then be contributing
to the shifted momentum, in contrast to the coordinate translation, 
and this could represent an overcounting of the terms containing the 
curvature tensor.  With the summation over the index $\rho$ included, 
the shifted momentum, denoted by $k^{\prime\prime}_\mu$ has an average 
squared value of
$\langle k^{\prime\prime 2} \rangle~=~k^2~\left[1~+~{{a^2}\over 2} 
\langle x^2 \rangle~+~{{a^4}\over {12}} \langle x^4 \rangle \right]$
Using either definition of the shifted momentum, the change is directly
related to the introduction of curvature in the manifold.

The momentum operator acts differently on wave functions by shifting the
argument, so that, for example, 
$e^{-i\alpha\cdot P} \psi(x)~=~\psi(x-\alpha)$.  This property would
be shared by the fields $X^\mu$ representing the coordinates of a string 
moving in a target space.  Choosing a fixed base point $X_0^\mu$ so that
$\Delta X^\mu~=~X^\mu~-~X_0^\mu$, one may define $\Delta X^A$ by
$\Delta X^\mu~=~~{{\partial x^\mu}\over {\partial X^A}} \Delta X^A$,
where $X^A$ represent standard embedding coordinates for the local 
anti-de Sitter geometry.  The action of the momentum operator $K$ is given by
$$(1-i \alpha\cdot K)\cdot \Delta X^A~=~\Delta X^A~-~\alpha^A
~-~{1\over 6}(R^A{}_{BCD}~+~R^A{}_{CBD}) \alpha^D \Delta X^B \Delta X^C
\eqno(81)$$ 
so that 
$$K^\prime_D~=~K_D~-~{i\over 6}(R^A{}_{BCD}~+~R^A{}_{CBD}) \Delta X^B
\Delta X^C {1\over {\Delta X^A}}
\eqno(82)$$
where there is no sum over the index A.  However, when squaring the
momentum, a special method of summing over the first index of the curvature
tensor and the index associated with ${1\over {\Delta X^A}}$ will be
used, and the sum shall be divided by 4, to average over all of the directions.
$$\eqalign{K^\prime_D K^{\prime D}~&=~K_D K^D~-~{i\over {12}}
\sum_A~(R^A{}_{BCD}~+~R^A{}_{CBD})
\Delta X^B \Delta X^C {1\over {\Delta X^A}} K^D
\cr
~~~~~~~~~
~&-~{1\over 4}\cdot {1\over {36}}~\sum_A~(R^A{}_{BCD}~+~R^A{}_{CBD}) 
\Delta X^B \Delta X^C
{1\over {\Delta X^A}} 
\cr
&~~~~~~~~~~~~~~~~~~~~~~\cdot (R_{AEF}{}^D~+~R_{AFE}{}^D) 
\Delta X^E \Delta X^F {1\over {\Delta X_A}}
\cr}
\eqno(83)$$
with an implied summation over the indices $A$,
or equivalently,
$$\eqalign{K^{\prime 2}=K^2-
&{{a^4}\over {144}}~\sum_A~\bigg[(\eta^A{}_C\eta_{BD}+
\eta^A{}_B\eta_{CD}-2\eta^A{}_D\eta_{CB})
\cr
&~~~~~~~~~~~~~~~~\cdot(\eta_{AF}\eta_E{}^D+\eta_{AE}\eta_F{}^D-
2\eta_A{}^D\eta_{EF})
 \cr
~&~~~~~~~~~~~~~~~~~~~\Delta X^B \Delta X^C \Delta X^E \Delta X^F
~\cdot~{1\over {\Delta X^A}}{1\over {\Delta X_A}}\bigg]
\cr
=K^2-&{{a^4}\over {144}}~\sum_A~(2\Delta X^A \Delta X_D~-
~2 \eta^A{}_D \Delta_B \Delta X^B)
             (2 \Delta X_A \Delta X^D~-~2\eta_A{}^D \Delta X_E \Delta X^E)
\cr             
&~~~~~~~~~~~~~~~~~~~~\cdot~{1\over {\Delta X^A}}{1\over {\Delta X_A}}
\cr}
\eqno(84)$$
with $\eta_{AD}~=~g_{\mu\nu}{{\partial x^\mu}\over {\partial X^A}}
{{\partial x^\nu}\over {\partial X^D}}$,
since
$$\eqalign{{1\over 4}~\sum_A~{i\over 3} a^2 (\eta_{AC} \eta_{BD}
~&+~\eta_{AB} \eta_{CD}~-~2 \eta_{AD} \eta_{BC}) \Delta X^B 
\Delta X^C~{1\over {\Delta X^A}} K^D
\cr
~&=~{{ia^2}\over 6}~\sum_A~
\left[\Delta X_A \Delta X_D~-~{{\eta_{AD}}\over a^2}\right]
{{\Delta X^A}\over {\Delta X_E \Delta X^E}} K^D~=~0
\cr}
\eqno(85)
$$
Using $\eta_{AB} \Delta X^A \Delta X^B~=~{1\over {a^2}}$, it follows that
$$\eqalign{\sum_{A,D}~\eta^A{}_D \Delta X^D {1\over {\Delta X^A}}
~=~\sum_{A,D}~\delta^\lambda{}_\sigma {{\partial {X^A}}\over 
{\partial {x^\lambda}}}{{\partial {x^\sigma}}\over {\partial {X^D}}}
&\Delta X^D {1\over {\Delta X^A}}
\cr 
~=~\sum_A~\delta^\lambda{}_\sigma {{\partial X^A}\over {\partial x^\lambda}}
{1\over {\Delta X^A}} &\Delta X^\sigma~=~4
\cr
\sum_{A,B}~\eta^A{}_D {1\over {\Delta X^A}} \eta_A{}^D 
{1\over {\Delta X_A}} \Delta X_B \Delta X^B
~=~\sum_{A,B}~&\delta^\lambda{}_\sigma \delta_\tau{}^\rho 
\delta^\sigma{}_\rho ~\left({{\partial {X^A}}\over {\partial x^\lambda}}
{{\partial x^\tau}\over {\partial X^A}} {1\over {\Delta X^A}}
{1\over {\Delta X_A}} \right)
\cr
&~~~~~~~~~~~~~~~~~~~~~~~~\Delta X_B \Delta X^B
\cr
~=~\sum_{A,B}~&{{\partial X^A}\over {\partial x^\tau}}
{{\partial x^\tau}\over {\partial X^A}}~{1\over {\Delta X^A}}
{1\over {\Delta X_A}}~\Delta X_B \Delta X^B
\cr} 
\eqno(86)
$$
The average value of the last sum in (86) is 16, so that the shift in the
squared momentum is $-{9\over {36}}a^2$.
Viewing the coordinate fields as a collection of scalar fields, 
the zero-point of the squared-mass scale would be $-{9\over 4}a^2$. 
Relative to this scale, the squared mass is shifted by ${9\over 4}a^2$, 
so that $m^{\prime 2}~=~m^2~+~{9\over 4}a^2$ and
$$K^{\prime 2}~+~m^{\prime 2}~=~K^2~+~m^2~+~ 2 a^2
\eqno(87)
$$
This suggests a connection with the bosonic string.  The Hamiltonian
for the closed bosonic string [17] is given by
$$H~=~{1\over 2}~\sum_n~[\alpha_{-n} \cdot \alpha_n~+~
{\tilde \alpha}_{-n}\cdot {\tilde \alpha}_n]
\eqno(88)
$$
where $\alpha_n^\mu$ and ${\tilde \alpha}_n^\mu$ are the coefficients in the
expansion of the target space coordinates $X^\mu$ that satifsy standard
operator commutation relations.  After imposing the operator condition
$L_0~=~{\tilde L}_0$ and normal ordering, it follows that the sum over 
transverse string oscillators can be assigned the value
$$\sum_{\mu=1}^{D-2} \sum_n~:\alpha_{-n}^\mu \alpha_{n\mu}:~+~(D-2)
~\sum_{n=1}^\infty~n
\eqno(89)$$
Upon use of the vanishing of the vacuum expectation value of the
operator product in the first sum, zeta-function regularization
of the second sum  gives a ground state contribution
of $-{{24}\over {12}}a^2$, after multiplying by the factor $a^2$,
which is dimensionally and numerically consistent with equation (87).
Since the energy-momentum of the string might be expected to curve the
space-time through which it propagates, it
would be useful to establish the relation between scattering in
a local region of constant curvature and string scattering. 
The coincidence might be explained by viewing the collection of 
bosonic string coordinates $\{X^\mu\}$ as part of a
string field $\Phi(\{X^\mu\})$.   Thus, although it is conceivable that the 
scattering of component fields should be considered in a curved local 
geometry, the ground state contribution to the string Hamiltonian 
appears to be compensated by the shift in $K^2+m^2$, indicating 
that the field-theoretic results might be included already in the entire 
string scattering calculations in flat space. 

A consideration of the various possiblities for the number of dimensions
shows that this coincidence between the shift in $K^2+m^2$ and the
magnitude of the residual contribution to the string Hamiltonian arising
after normal ordering can only be achieved in four dimensions.  The
momentum shift can also be calculated in the space of positive constant
curvature, de Sitter space.  After replacing $a^2$ by $-a^2$ in
equations (79), (80) and (84), it can be seen that the shift in the 
Hamiltonian can be represented as $K^{\prime\prime 2}+m^{\prime\prime 2}
~=~K^2+m^2-2a^2$.  The shift of $-2a^2$ has also been obtained by regularizing
the sum over transverse string oscillators in (89).  Given this alternative
representation of the string Hamiltonian, it might be conjectured that the
target space geometry would be locally altered to de Sitter space.
Scattering amplitudes in geometries which are locally de Sitter space and
globally flat have not yet been computed, although a conformal transformation
of the local target space metric to flat space could be used.  Two possible
models of string propagation in the interaction region therefore exist.
First, the energy-momentum associated with the string might curve the
background geometry so that it is locally anti-de Sitter space and then the
expression for the string Hamiltonian receives the extra contribution
(87) which would cancel the normal-ordering effect in (89), consistent
with the conventional formulation of scattering amplitudes in flat space.
The other alternative is that the overall effect of the string 
energy-momentum is a momentum shift which reveals a geometry with a 
locally de Sitter metric.

The latter possibility probably can be eliminated because supersymmetry 
requires a flat or anti-de Sitter background, whereas theories with de Sitter
supersymmetry contain negative-norm states [6].  This is confirmed by a
study of string theories in curved space-times, based on coset conformal field
theories [18].  Although the bosonic and superstring theories are typically 
viewed as independent, they can be combined in the heterotic string theory, 
and since this theory is connected by duality to the other string theories, 
this indicates that the first description of string propagation in a globally 
flat space-time is more appropriate.

A calculation of more physical relevance is the scattering of the
superstring in ten dimensions.  For type II theories, the Hamiltonian is
$$H~=~{1\over {2\pi p^+}}~\int_0^\pi~d\sigma~[\pi^2 (P_\tau^i)^2~+~
(X^{i\prime})^2~-~iS^1 S^{1\prime}~+~iS^2 S^{2\prime}]
\eqno(90)$$
where $S^{1a}$ and $S^{2a}$ are one-component Majorana-Weyl world-sheet
spinors describing right-moving and left-moving degrees of freedom [19].
Expanding
$$\eqalign{X^\mu~&=~x^\mu~+~p^\mu \tau
~+~i\sum_{n\ne 0}~{1\over n}~\alpha_n^\mu~e^{-in \tau}~cos~n \sigma
\cr
\psi_+^\mu~&=~\sum_r~b_r^\mu~e^{-2ir(\tau-\sigma)}~~~~~~~(NS)
\cr
\psi_-^\mu~&=~\sum_r~{\tilde b}_r^\mu~e^{-2ir(\tau+\sigma)}~~~~~~~(NS)
\cr
\psi_+^\mu~&=~\sum_n~{d_n^\mu}~e^{-2in(\tau - \sigma)}~~~~~~~(R)
\cr
\psi_-^\mu~&=~\sum_n~{\tilde d}_n^\mu~e^{-2in(\tau+\sigma)}~~~~~~~(R)
\cr}
\eqno(91)
$$
where
$$\eqalign{[\alpha_m^\mu, \alpha_n^\nu]~&=~[{\bar \alpha}_m^\mu,
{\bar \alpha}_n^\mu]~=~0
\cr
\{b_r^\mu,~b_s^\nu\}~&=~\eta^{\mu\nu}~\delta_{r,-s}~~~~~~~(NS)
\cr
\{d_n^\mu,~d_m^\nu\}~&=~\eta^{\mu\nu} \delta_{n,-m}~~~~~~~(R)
\cr}
\eqno(92)
$$
The Virasoro operators are
$$\eqalign{L_m~&=~L_m^{(\alpha)}~+~L_m^{(b)}~~~~~~~~~~(NS)
\cr
L_m~&=~L_m^{(\alpha)}~+~L_m^{(d)}~~~~~~~~~~(R)
\cr}
\eqno(93)
$$
so that
$$\eqalign{L_0^{(\alpha)}~&=~{1\over 2}~\sum_{n=-\infty}^\infty~:\alpha_{-n}
\cdot \alpha_n:
\cr
L_0^{(b)}~&=~{1\over 2}~\sum_r~r~:b_{-r} \cdot b_r:
~~~~~~~~half-integrally~moded
\cr
L_0^{(d)}~&=~{1\over 2}~\sum_n~n~:d_{-n} \cdot d_n:
~~~~~~~~integrally~moded
\cr}
\eqno(94)
$$
The normal ordering constant from the physical bosonic coordinate is
${1\over {24}}$ [$\epsilon_B^+~=~-{1\over {24}}$], 
while the normal ordering constant from a half-integrally
moded fermionic coordinate is ${1\over {48}}$ [$\epsilon_F^-~=~-{1\over {48}}$]
and from an integrally moded
fermionic coordinate is $-{1\over {24}}$, [$\epsilon_F^+~=~{1\over {24}}$], 
so that the bosonic contribution to the Hamiltonian is $-{8\over {12}}$.  
Now consider superstring scattering in a background where four of the 
dimensions locally represent anti-de Sitter space and six dimensions are 
compactified with radius of curvature significantly greater than 
the dimensions of the string.  The dominant contribution to the shift 
in the momentum arises from four-dimensional anti-de Sitter 
space.  Viewing 
$\Delta X^\mu~=~{{\partial x^\mu}\over {\partial X^A}} \Delta X^A$ 
as a vector field on this space, and noting that spin-s 
fields satisfy the equation 
$$[C_2+[\omega_0(\omega_0-3)~+~s(s+1)]a^2~]\chi_s~=~0
\eqno(95)$$        
where $C_2~=~{1\over 2}J^{AB}J_{AB}$ is the second Casimir invariant, 
it follows that the squared mass is shifted by ${1\over 4} a^2$ for each
component of the vector field $\Delta X^\nu$.  Multiplying  
this result by 4, associated with the four dimensions, gives $2a^2$.
The shift in the vector field $\Delta X^\nu$ is given by a Lie derivative 
in the direction of the momentum field, or equivalently the commutator of
the two fields.  This has already been anticipated in equation (74).
From equation (80), and conversion of the product of $k^\mu$ and $x^\mu$,
the position coordinate also denoted by $\Delta X^\mu$, 
through the replacement of the flat-space momentum operator by 
$-i{\partial\over {\partial x^\mu}}$, it can be shown the average value of the
squared momentum becomes
$\langle k^{\prime 2}\rangle~=~k^2~-~{{a^2}\over 8}
\langle {1\over {x^3}}{d\over {dx}}x^3 {d\over {dx}}(x^2) \rangle
~-~{{a^4}\over {192}}\langle {1\over {x^3}}{d\over {dx}} x^3 
{d\over {dx}}(x^4) \rangle
~=~k^2~-~a^2~-~{{a^4}\over 8} \langle x^2 \rangle$.
The distribution of measurements of the variable $x~=~(x_\mu x^\mu)^{1\over 2}$
will be a normal distribution $N(\mu, \sigma^2)$.  Given the dimensions of
the anti-de Sitter geometry, the mean value of $x$ can be set equal to 
${1\over a}$, while the variance may be chosen initially so that the 
expectation value of $x^2$ produces the required shift in the 
squared momentum, $k^{\prime 2}~=~k^2~-~{4\over 3}a^2$.  This can be
achieved with an expectation value $\langle x^2 \rangle~=~{8\over {3a^2}}$
and variance $\sigma^2~=~{5\over {3a^2}}$.  The underlying reasons
for the occurrence of the normal distribution $N({1\over a}, {5\over {3a^2}})$
have yet to be determined, although the number of variables $x^\mu$ and 
any correlations between these coordinates would affect the variance
$\sum_\mu Var(x^\mu)~+~2~\sum_{\mu < \nu} Cov(x^\mu, x^\nu)$,
as the covariance $Cov(x^\mu, x^\nu)$ measures the correlation between
$x^\mu$ and $x^\nu$.  Given this momentum shift, that the $K^2~+~m^2$ is 
shifted by ${2\over 3} a^2$, compensating the bosonic ground state 
contribution to the Hamiltonian calculated earlier.
The fermionic contribution can be determined in the identical manner 
using the normal ordering of the corresponding oscillators.  The difference 
in the interpretation of the coordinate fields $\{\Delta X^\mu\}$ of
the bosonic string and superstring might be traced to the inclusion of
the fermions, which necessarily must combine to produce a worldsheet
vector field representing the projection of a target-space vector field
mediating the interactions.  

Although the definition mentioned above is motivated by physical considerations
such as 
position-momentum commutation relations, it is not necessarily
the only one to use in a quantum theory.  Other possibilities include
the transform space variables such as $\lambda$ in equation (23),
the flat-space version of momentum 
used in the adiabatic expansion of Green functions [20][21][22],
and even the generator $aJ_{\mu 4},~\mu~=~0,1,2,3$, which does not form
a commutative subalgebra but does have the property that it is conserved
along the particle's worldline.  The last expression for the momentum is
$$P_\mu~=~p_\mu~-~{{ia}\over m}~p^\nu~J_{\nu\mu}
\eqno(96)
$$
for a particle of mass $m$ [22], and the square is
$$\eqalign{P_\mu~P^\mu~&=~p_\mu~p^\mu~-
~2{{ia}\over m}~p^\mu~p^\nu~J_{\nu\mu}~-~
{{a^2}\over {m^2}}~p^\nu~p^\sigma~J_{\nu\mu}~J_\sigma{}^\mu
\cr
~&=~p_\mu~p^\mu~-~{{a^2}\over {m^2}}~p^\nu~p^\sigma~J_{\nu\mu}
~J_\sigma{}^\mu
\cr}
\eqno(97)
$$
where $p_\mu$ is a flat-version of the momentum defined to be 
$m{\tilde x}_\mu$, with the four-vectors ${\tilde x}_\mu$ satisfying 
${\tilde x}_0~>~0$ and ${\tilde x}^\mu~{\tilde x}_\mu~=~1$ [23].  
In contrast to the other definitions 
of momentum, these quantities do not represent dual space variables 
with respect to position coordinates and therefore cannot 
be used directly in the calculation of loop diagrams and renormalized 
energy-momentum tensors.  From the calculations of the shifts 
in the squared momentum and squared mass in anti-de 
Sitter space, relations between the different definitions of the 
momenta have been deduced.  It would be of interest to determine whether 
any of these connections are maintained for general curved spaces.  The 
conclusions may be relevant for recent work in curved space quantum field 
theory regarding the improved calculation of propagators [24] 
for fields of spin 0, ${1\over 2}$ and 1, which may be useful for 
determining particle-production rates, and localized renormalization 
theory [25].

\vfill
\eject
 
\centerline{\bf Acknowledgements}

I would like to thank Dr Achim Kempf for first elucidating the 
effect of curvature on position-momentum commutation
relations and Dr Hugh Luckock for a useful discussion about momentum shifts
and string theory.  This research was initiated in the University of Cambridge
and completed at the University of Sydney.  A Royal Society Study
Visit Award is gratefully acknowledged.

\vfill\eject

\centerline{\bf Appendix}

It is possible to find the analogue of hyperplanes in homogeneous spaces
$X~=~G/H$, for which there exists an isometry reversing geodesics,
or equivalently a translation-invariant Riemann tensor [5].

Any semi-simple Lie group may be decomposed into a compact and non-compact
part.  The Lie algebra may be written as a direct sum of the vector spaces
associated with the compact and non-compact generators: ${\cal G}~=~{\cal H}
~\oplus~{\cal P}$, the Cartan decomposition.  
Furthermore, there exists an isometry $\Theta$ which
maps ${\cal H}$ into ${\cal H}$ and ${\cal P}$ into $-{\cal P}$.
The classification of locally symmetric spaces can then be reduced to the
listing of all compact Lie algebras together with the isometry $\Theta$.

A Cartan symmetric space is of the form $X~=~G/H$ where H is the maximal
compact subgroup.  There exists another decomposition,
known as the Iwasawa decomposition, which is more useful for our purposes.
Let $u~=~{\cal H} + {\cal P}_0$ where ${\cal P}_0~=~i{\cal P}$ is a compact 
form of ${\cal G}$, and let ${\cal A}$ be a maximal abelian subalgebra
of ${\cal P}_0$.  Consider the subspace
$${\cal G}_\alpha~=~\{{\bf g}\in {\cal G }\vert [{\bf g}, {\bf a}]
~=~\alpha({\bf a}){\bf g}~for~all~
{\bf a} \in {\cal A}\}
\eqno(98)
$$
where $\alpha({\bf a})$ is called the restricted root.  Thus, 
${\cal G}~=~\sum_\alpha~{\cal G}_\alpha~+~{\cal G}_0$ 
and ${\cal A} \subset {\cal G}_0$.  The subset of generators $\{{\bf a}\}$
in ${\cal A}$
for which all the restricted roots are non-zero has a connected component
${\cal A}^+$, the Weyl chamber.

A root $\alpha$ is positive if $\alpha({\bf a}) > 0$ for all 
${\bf a}\in {\cal A^+}$.
Now let ${\cal N}~=~\sum_{\alpha>0} {\cal G}_\alpha$.  Then, the Iwasawa 
decomposition of the Lie algebra ${\cal G}$ is
$${\cal G}~=~{\cal H}~\oplus~{\cal A}~\oplus~{\cal N}
\eqno(99)$$
and exponentiating this expression, $g~=~han$, $g\in G$, $h\in H$, $a\in A$
and $n\in N$.  The nilpotency of N as subgroup of G follows from the
existence of $n$ such that $N^{(n)}~=~0$, with the definition
$$N^{(0)}~=~N,~~~N^{(1)}~=~[N,N],~~~N^{(2)}~=~[N,[N,N]],...
\eqno(100)$$

This decomposition can be used to define generalized plane waves or
horocycles in $X$.  Consider a point $o$, whose stability subroup is $H$,
to be the origin of $X$. The simplest horocycle is the orbit 
$\xi_0~=~N\cdot o$.  All horocycles in $X$ can be
written as $\xi~=~gNg^{-1}{\tilde g}\cdot o$, 
$g,{\tilde g}\in G$.  Since ${\tilde g}^{-1}g~=~han$,
$$\xi~=~{\tilde g}({\tilde g}^{-1}g)N({\tilde g}^{-1}g)^{-1}\cdot o
~=~{\tilde g}han N n^{-1}a^{-1}h^{-1}\cdot o
~=~{\tilde g}haNa^{-1}\cdot o~=~{\tilde g}h N\cdot o
\eqno(101)
$$
since $aNa^{-1} \subset N$, and therefore,
$$\xi~=~{\tilde g}h\cdot \xi_0~=~[k({\tilde g}h)]
[a({\tilde g}h)][n({\tilde g}h)]N\cdot o~=~[k({\tilde g}h)][a({\tilde g}h)]
N\cdot o
~\equiv~{\tilde h}{\tilde a}\cdot \xi_0
\eqno(102)
$$
Every horocycle may be written in the form $\xi~=~ha\cdot \xi_0$ where
$\xi_0$ is the fundamental horocycle.

Since $a$ is an element of a group, it can be expressed as $exp~r$ where
$r\in {\cal A}$ represents the distance from the origin to the horocycle
$\xi$.  However, a horocycle is not specified by a particular choice of
$h$ and $a$.  In fact, if we define the centalizer of ${\cal A}$ in $H$ by
$M~=~\{h\in H \vert Ad(h){\bf a} ~=~{\bf a}~for~all~{\bf a}\in {\cal A} \}$, 
then
$h$ and $h^\prime$ give the same horocycle if they belong to the
same coset $H/M$.  If $h~=~h^\prime m$,
$$\xi~=~haN\cdot o~=~h^\prime m a N\cdot o~=~ h^\prime a N m\cdot o
~=~ h^\prime a \cdot \xi_0
\eqno(103)
$$
since $M$ stabilizes $AN$.  Finally, defining the boundary of the symmetric
space to be $B~=~H/M$, only one horocycle passes through $x \in X$ with
normal $b\in B$.

Although this formalism cannot be adapted to anti-de Sitter space,
SO(3,2)/SO(3,1), with a non-compact stability group, $H^4~=~SO(4,1)/SO(4)$
is a Cartan symmetric space.  Results obtained in $H^4$ may then
be analytically continued back to anti-de Sitter space, by analogy with
a Wick rotation from a Euclidean field theory to a Lorentzian field theory.

Since the Lie algebra so(p,q) is given by
$$\left\{\left(\matrix{X_1&X_2
                        \cr
                      X_2^t&X_3
                          \cr}
                            \right)~\bigg|~X_1=-X_1^t,~X_3=-X_3^t\right\}
\eqno(104)
$$                                            
Since ${\cal G}~=~so(4,1)$, ${\cal K}~=~so(4)$ and 
$so(4,1)~=~so(4)~+~{\cal P}$ and the maximum abelian subalgebra ${\cal A}$ in 
${\cal P}$ is 
$${\Bbb R} \left(\matrix{0&0&0&0&1
                           \cr
                         0&0&0&0&0
                           \cr
                         0&0&0&0&0
                            \cr
                         0&0&0&0&0
                             \cr
                         1&0&0&0&0
                             \cr}
                                  \right)
\eqno(105)
$$
from the commutation relations of ${\cal P}$ with an arbitrary element of the 
Lie algebra ${\cal G}$, it can be shown that the roots $\{\alpha\}$ assume the
values 1 or ${-1}$.  The decomposition of the so(4,1) is then
$$\eqalign{{\cal G}~&=~{\cal G}_{+1}~+~ {\cal G}_{-1}~+~{\cal G}_0
\cr
~=~{\Bbb R}~&\left(\matrix{0&-1&0&0&0
                        \cr
                       1&0&0&0&-1
                         \cr
                       0&0&0&0&0
                          \cr
                       0&0&0&0&0
                          \cr
                       0&-1&0&0&0
                           \cr}
                                \right)
+{\Bbb R}~\left(\matrix{0&0&-1&0&0
                            \cr
                         0&0&0&0&0
                             \cr
                         1&0&0&0&-1
                            \cr
                         0&0&0&0&0
                             \cr
                         0&0&-1&0&0
                              \cr}
                                \right)
+{\Bbb R}~\left(\matrix{0&0&0&-1&0
                            \cr
                         0&0&0&0&0
                             \cr
                         0&0&0&0&0
                             \cr
                         1&0&0&0&-1
                              \cr
                         0&0&0&-1&0
                              \cr}
                                \right)
\cr
~+~{\Bbb R}~&\left(\matrix{0&-1&0&0&0
                              \cr
                          1&0&0&0&1
                              \cr
                          0&0&0&0&0
                              \cr
                          0&0&0&0&0
                              \cr
                          0&1&0&0&0
                              \cr}
                               \right)
+{\Bbb R}~\left(\matrix{0&0&-1&0&0
                              \cr
                         0&0&0&0&0
                              \cr
                         1&0&0&0&1
                             \cr
                         0&0&0&0&0
                             \cr
                         0&0&1&0&0
                              \cr}
                                 \right)
+{\Bbb R}~\left(\matrix{0&0&0&-1&0
                            \cr
                          0&0&0&0&0
                             \cr
                          0&0&0&0&0
                             \cr
                          1&0&0&0&1
                              \cr
                          0&0&0&1&0
                               \cr}
                                  \right)
\cr
~+~{\Bbb R}~&\left(\matrix{0&0&0&0&0
                             \cr
                          0&0&-1&0&0
                              \cr
                          0&1&0&0&0
                              \cr
                          0&0&0&0&0
                               \cr
                          0&0&0&0&0
                              \cr}
                                  \right)
+{\Bbb R}~\left(\matrix{0&0&0&0&0
                             \cr
                          0&0&0&-1&0
                             \cr
                          0&0&0&0&0
                               \cr
                          0&1&0&0&0
                              \cr
                          0&0&0&0&0
                               \cr}
                                  \right)
+{\Bbb R}~\left(\matrix{0&0&0&0&0
                               \cr
                          0&0&0&0&0
                                \cr
                          0&0&0&-1&0
                               \cr
                          0&0&-1&0&0
                                \cr
                          0&0&0&0&0
                                \cr}
                                  \right)
\cr
&~+~{\Bbb R}~\left(\matrix{0&0&0&0&1
                            \cr
                         0&0&0&0&0
                            \cr
                         0&0&0&0&0
                             \cr
                         0&0&0&0&0
                             \cr
                         1&0&0&0&0
                             \cr}
                                 \right)
\cr}                         
\eqno(106)
$$
The nilpotent subalgebra is ${\cal N}~=~{\cal G}_{+1}$ and
$$\eqalign{N_1^2~&=~N_2^2~=~N_3^2~=~\left(\matrix{-1&0&0&0&1
                                                  \cr
                                                 0&0&0&0&0
                                                    \cr
                                                 0&0&0&0&0
                                                     \cr
                                                 0&0&0&0&0
                                                    \cr
                                                 -1&0&0&0&1
                                                     \cr}
                                                      \right)
\cr
N_1^3~&=~N_2^3~=~N_3^3~=~0
\cr
N_1 N_2~&=~N_1 N_3~=~N_2 N_3~=~0
\cr}
\eqno(107)
$$
The general element of the nilpotent group is given by
$$\eqalign{exp(n_1N_1+&n_2N_2+n_3N_3)
\cr
~&=~1~+~n_1N_1~+~n_2N_2~+~n_3N_3
                                ~+~{1\over 2}(n_1^2N_1^2+n_2^2N_2^2
                                                    +n_3^2N_3^2)
\cr
~=~&\left(\matrix{1-{1\over 2}(n_1^2+n_2^2+n_3^2)&-n_1&-n_2&
-n_3&{1\over 2}(n_1^2+n_2^2+n_3^2)
\cr
n_1&1&0&0&-n_1
\cr
n_2&0&1&0&-n_2
\cr
n_3&0&0&1&-n_3
\cr
-{1\over 2}(n_1^2+n_2^2+n_3^2)&-n_1&-n_2&-n_3^2&1+{1\over 2}(n_1^2+n_2^2+n_3^2)
\cr}
\right)
\cr}
\eqno(108)
$$
Since $H$ rotates the first four coordinates, the point $o~=~(0,0,0,0,1)$
is chosen to be the origin and the fundamental horocycle is given by
$$N\cdot o~=~\left(\matrix{-{1\over 2}(n_1^2+n_2^2+n_3^2)&
                                            \cr        
                                            -n_1&
                                            \cr
                                            -n_2&
                                            \cr
                                            -n_3&
                                            \cr
                            1+{1\over 2}(n_1^2+n_2^2+n_3^2)&
                                       \cr}
                                          \right)
~~~~~~~~n_1,~n_2,~n_3~\in~{\Bbb R}
\eqno(109)
$$
To determine the distance from the origin to the horocycle passing through
$x$ with normal $b~=~hM$, it may be noted that if $x\in \xi~=~haN\cdot o$,
then $an\cdot o~=~h^{-1}\cdot x$ for some $n \in N$.  
Using $\sum_{i=0}^3(h^{-1}\cdot x)_i^2~=~\sum_{i=0}^3 x_i^2$, 
it can be shown that the matrix equation

$$\left(\matrix{cosh~r&0&0&0&sinh~r
                      \cr
                   0&1&0&0&0
                       \cr
                  0&0&1&0&0
                      \cr
                  0&0&0&1&0
                      \cr
                  sinh~r&0&0&0&cosh~r
                       \cr}
                          \right)
                         \left(\matrix{-{1\over 2}(n_1^2+n_2^2+n_3^2)&   
                                                    \cr
                                                 -n_1&
                                                  \cr
                                                 -n_2&
                                                  \cr
                                                 -n_3&
                                                  \cr                    
                                               1+{1\over 2}(n_1^2+n_2^2+n_3^2)&
                                                  \cr}\right)
                                     ~=~\left(\matrix{(h^{-1}\cdot x)_0&
                                                             \cr
                                                      (h^{-1}\cdot x)_1&
                                                             \cr
                                                      (h^{-1}\cdot x)_2&
                                                             \cr
                                                      (h^{-1}\cdot x)_3&
                                                             \cr
                                                             x_4& 
                                                             \cr}
                                                             \right)
\eqno(110)
$$        
has solutions $n_i~=~-(h^{-1}\cdot x)_i$ and 
$$\eqalign{cosh~r~&=~{1\over 2}
\left[{1\over {x_4-(h^{-1}\cdot x)_0}}~+~x_4-(h^{-1}\cdot x)_0 \right]
\cr
r~&=~ln[x_4-(h^{-1}\cdot x)_0]~=~ln[h^{-1}\cdot x,\xi]~=~ln [x, h\xi]
~~~~~~~\xi~=~(1,0,0,0,1)
\cr}
\eqno(111)
$$
This distance is independent of $(h^{-1}\cdot x)_i,~i=1,2,3$ as the rotation
of these three coordinates corresponds to the SO(3) subgroup $M$ which
stabilizes ${\cal A}$.  Indeed, since ${\cal A}~=~J_{04}$, $M$ is generated
by $J_{12},~J_{13},~J_{23}$ since these are the only compact generators 
which commute with $J_{04}$.   Consequently, the boundary is B=SO(4)/SO(3).
 
The group-theoretical definition of horocycles, or equivalently horospheres,
agrees with the geometric one.  
The geometric definition is given as follows: consider a point $x_0$ in
$H^4$ and draw all geodesics through it.  A sphere of radius r with center
at $x_0$ is the set of points which are a distance $r$ from $x_0$ on the
geodesics, $\{exp~rX \vert X \in T_{x_0}(M)\}$. A horosphere 
is a sphere with center at infinity that still passes through a specified
point in $H^4$. 

Lobachevskii space can be identified with the set of lines through the
origin with the metric
$$cosh~r~=~{{[x,y]}\over {[x,x][y,y]}}~~~~~x,~y~are~arbitrary~points~on~two~
different~lines
\eqno(112)
$$
Instead of lines, a representative point can be chosen from each line,
for example the intersection with $[x,x]~=~1$.  Then, $cosh~r~=~[x,y]$.
A point at infinity may be represented by a null vector on the cone,
$-\xi_0^2-\xi_1^2-\xi_2^2-\xi_3^2+\xi_4^2~=~0$.  Thus, the equation of the
horosphere is $[x,\xi]~=~const.$ and by rescaling $\xi$, one obtains
$[x, \xi]~=~1$, defining a horosphere of the first kind.
In imaginary Lobachevskii space (or de Sitter space in 4 dimensions), 
horospheres of both the first and second kind exist [26], where the latter type
is given by the equation $[x, \xi]=0$.  

Let $\xi$ be a particular null vector and $a$ be a point on the hyperboloid.  
Then, it can be shown that $ln~[a,\xi]$ equals the distance from $a$ to 
the horosphere $[x,\xi]~=~1$, defined to be the distance to the 
closest point on that horosphere.  Since the scalar product is invariant 
under the action of the group,
$$\eqalign{[a,\xi]~&=~[ga,g\xi]~=~[a^\prime, \xi^\prime]~=~t
\cr
a^\prime~&=~(0,0,0,0,1)~~~~~~\xi^\prime~=~(t,0,0,0,t)
\cr}
\eqno(113)
$$                                                                        
The distance from $a$ to the horosphere $[x,\xi]~=~1$ equals the distance
from $a^\prime$ to the horosphere $[x,\xi^\prime]~=tx_0~+~tx_4~=~1$.  
The distance from $a^\prime$ to $[x,\xi^\prime]~=~1$ is the radius of the
sphere centered at $a^\prime$ and tangent to $[x,\xi^\prime]~=~1$, and
every sphere is the intersection of the $[x,x]~=~1$ hyperboloid with the
hyperplane $x_4~=~constant$.  The sphere is tangent with $[x,\xi^\prime]~=~1$
at the point $y$, where $y_4$ assumes its minimum value.  Since
$y_4^2~-~y_0^2~=~1~+~y_1^2~+~y_2^2~+~y_3^2$, minimizing $y_4$ implies
that $y_1~=~y_2~=~y_3~=~0$, and since $y_0~=~-t^{-1}~+~y_4$, 
${y_4}_{min}~=~{1\over 2}(t~+~t^{-1})$.  Again the distance is given by
$$\eqalign{cosh~r~&=~[a^\prime,y]~=~{1\over 2}(t~+~t^{-1})
\cr
r~&=~ln~t~=~ln~[a,\xi]
\cr}
\eqno(114)
$$                                                    

Thus, choosing $\xi~=~(1,0,0,0,1)$, it follows that the distance from 
$h^{-1}\cdot x$ to the horocycle $[z,\xi]~=~1$ equals
$ln~[h^{-1}\cdot x,\xi]~=~ln~[x,h\xi]$, which is the distance from
$x$ to $[z,h\xi]~=~1$.  Since $h$ does not change the last coordinate of
$\xi$, $(0,0,0,0,1)$ is still a point on this horocycle.

The geometrical definition of $r(x,b)$ is therefore the
distance from $x$ to the horocycle passing through (0,0,0,0,1) with
normal $b~=~hM$, whereas the group-theoretical definition of $r(x,b)$
is the distance from $(0,0,0,0,1)$ to the horocycle passing through $x$
with normal $b$.

The equivalence between these two definitions is analogous to the equality
in Euclidean space between the distance from the origin $o$ 
to the hyperplane passing through $x$ with normal $w$ 
and the distance from $x$ to the hyperplane through the origin 
$(z,w)~=~0$ (Fig. 3).

The equality of the two distances for horocycles will now be shown.  Let
$y$ be the point on the horocycle through $(0,0,0,0,1)$ closest to $x$ and 
$z$ be the point on the horocycle through $x$ closest to $(0,0,0,0,1)$.
The isometry which maps $(0,0,0,0,1)$ into $z$ also maps the horocycles
into each other.  This means that there must be a point $y^\prime$
on the horocycle at the same distance from $x$ as the distance 
between $(0,0,0,0,1)$ and $z$.  The point $y$ is obtained by drawing a 
geodesic intersecting $[x,\xi]~=~1$ orthogonally.  If the geodesic
is continued to the boundary, it will reach the same boundary
point as the geodesic from $(0,0,0,0,1)$ to $z$.  As there is an isometry
mapping the two geodesics into each other, while keeping the horocycles
fixed, the distances must be the same.  Essentially, the geodesics and
the horocycles form an orthogonal coordinate system.

\vbox{
\epsfysize=2.5in
\epsfxsize=4in
\centerline{\epsfbox{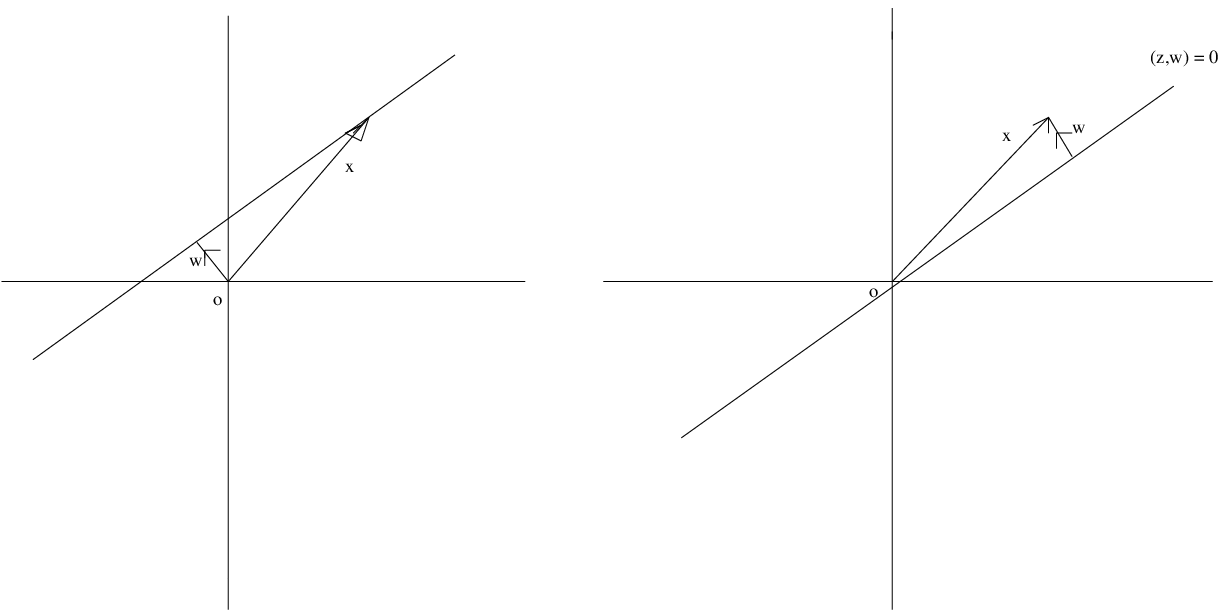}}
\vskip 0.2in
\noindent{\bf Fig. 3  Equivalence between the distance from the origin to the
hyperplane passing through x with normal w and the distance from x to the
hyperplane through the origin.}}

Since $r(x,b)~=~ln~[x,\xi]$, generalized plane waves in $H^4$ have the form
$[x,\xi]^\sigma$.  It can be verified that $[x,\xi]^\sigma$ is an eigenfunction
of the Laplacian with eigenvalue $\sigma(\sigma+3)$.
Using coordinates $\eta,~{\hat x},~{\hat y},~{\hat z}$ defined by
$$\eqalign{X_0~&=~{{\eta^{-1}~-~\eta}\over 2}~+~{1\over 2}\eta^{-1}
({\hat x}^2+{\hat y}^2+{\hat z}^2)
\cr
X_1~&=~\eta^{-1}{\hat x}~~~~~X_2~=~\eta^{-1}{\hat y}~~~~~X_3~=~\eta^{-1}
{\hat z}
\cr
X_4~&=~{{\eta+\eta^{-1}}\over 2}~+~{1\over 2}\eta^{-1}({\hat x}^2
+{\hat y}^2+{\hat z}^2)
\cr}
\eqno(115)
$$
so that 
$ds^2~=~{1\over {\eta^2}}~[d\eta^2~+~d{\hat x}^2~+~d{\hat y}^2~+~d{\hat z}^2]$.
Then,
$$\eqalign{[x,\xi]^\sigma~&=~-{1\over 2}\eta^{-1}\xi_0 
[1~-~\eta^2~-~({\hat x}^2~+~{\hat y}^2~+~{\hat z}^2)]
~-~\eta^{-1}\xi_1 x~-~\eta^{-1}\xi_2 y~-~\eta^{-1}\xi_3 z
\cr
&~~~~~+~{1\over 2}\eta^{-1}\xi_4 [1~+~\eta^2~+~{\hat x}^2~+~{\hat y}^2
~+~{\hat z}^2]
\cr}
\eqno(116)
$$
Since
$$\square~=~{1\over {\sqrt g}}\partial_\mu ({\sqrt g} g^{\mu\nu}\partial_\nu)
~=~\eta^2{{\partial^2}\over {\partial\eta^2}}~-~2\eta 
{\partial\over {\partial \eta}}~+~\eta^2\left
({{\partial^2}\over {\partial {\hat x}^2}}~+~
{{\partial^2}\over {\partial {\hat y}^2}}
~+~{{\partial^2}\over {\partial {\hat z}^2}}\right)
\eqno(117)
$$

$$\eqalign{\square~[x,\xi]^\sigma~=&~\eta^2~\sigma(\sigma-1)~[x,\xi]^{\sigma-2}
\{{1\over 2}\xi_0[\eta^{-2}+1+\eta^{-2}({\hat x}^2+{\hat y}^2
+{\hat z}^2)]
\cr
&~+~\eta^{-2}(\xi_1 {\hat x}+\xi_2 {\hat y}+\xi_3 {\hat z})
-~{1\over 2}\xi_4~[\eta^{-2}
-1+\eta^{-2}({\hat x}^2+{\hat y}^2+{\hat z}^2)]\}^2
\cr
&~-~\eta^2 \sigma [x,\xi]^{\sigma-1}
\{\xi_0 \eta^{-3}[1-({\hat x}^2+{\hat y}^2+{\hat z}^2)]
\cr
&~+~2\eta^{-3}(\xi_1 {\hat x}+\xi_2 {\hat y}~+\xi_3 {\hat z})
~-~\xi_4 \eta^{-3}~[1+{\hat x}^2+{\hat y}^2+{\hat z}^2]\}
\cr
&~-~\eta\sigma~[x,\xi]^{\sigma-1} \{\xi_0~[\eta^{-2}+1-\eta^{-2}
({\hat x}^2+{\hat y}^2+{\hat z}^2)]
\cr
&~+~
2\eta^{-2}(\xi_1 {\hat x}+\xi_2 {\hat y}+\xi_3 {\hat z})
~-~\xi_4~[\eta^{-2}-1+\eta^{-2}({\hat x}^2+{\hat y}^2+{\hat z}^2)]\}
\cr
&~+\eta^2\sigma(\sigma-1)~[x,\xi]^{\sigma-2}\eta^{-2}
~[(\xi_0+\xi_4){\hat x}-\xi_1]^2
+\eta^2\sigma~[x,\xi]^{\sigma-1}\eta^{-1}(\xi_0+\xi_4)
\cr
&~+\eta^2~\sigma(\sigma-1)~[x,\xi]^{\sigma-2}\eta^{-2}
[(\xi_0+\xi_4){\hat y}-\xi_2]^2+\eta^2 \sigma [x,\xi]^{\sigma-1}\eta^{-1}
(\xi_0+\xi_4)
\cr
&~+\eta^2~\sigma(\sigma-1)~[x,\xi]^{\sigma-2}\eta^{-2}
[(\xi_0+\xi_4){\hat z}-\xi_3]^2+\eta^2 \sigma[x,\xi]^{\sigma-1}\eta^{-1}
(\xi_0+\xi_4)
\cr
~=&~\eta^{-2}~\sigma(\sigma-1)~[x,\xi]^{\sigma-2}
[\{{1\over 2}\xi_0~[1+\eta^2-({\hat x}^2+{\hat y}^2+{\hat z}^2)]~+~
(\xi_1 {\hat x}+\xi_2 {\hat y}+\xi_3 {\hat z}) 
\cr                    
&~-~{1\over 2}\xi_4~[1+({\hat x}^2+{\hat y}^2+{\hat z}^2)]
~-~{1\over 2}(\xi_0+\xi_4) \eta^2 \}^2
       +\eta^2\{(\xi_0+\xi_4){\hat x}-\xi_1\}^2
\cr
&~+~\eta^2\{(\xi_0+\xi_4){\hat y}-\xi_2\}^2~+
~\eta^2\{(\xi_0+\xi_4){\hat z}~-~\xi_3\}^2]
+4\sigma~[x,\xi]^\sigma
\cr
~=&~\sigma(\sigma-1)~[x,\xi]^{\sigma-2}~[x,\xi]^2~+~4\sigma[x,\xi]^\sigma
~=~\sigma(\sigma+3)~[x,\xi]^\sigma
\cr}
\eqno(118)
$$  
using the property that $-\xi_0^2~-~\xi_1^2~-~\xi_2^2~-~\xi_3^2~+~\xi_4^2~=~0$.

The hyperboloid $H^4$ is a two-point homogeneous space, so that any two
pairs of points separated by the same distance may be mapped into each other
by an isometry.  As any two-point homogeneous space has rank 1, the vector
space of G-invariant differential operators, $D(G/K)$ has one generator,
the d'Alembertian. The Dirac operator, for example, is not an invariant
differential operator, and the generalized plane waves are not eigenfunctions,
in contrast to flat space, where 
$(i\gamma\cdot\partial-m)e^{-ip\cdot x}~=~(\gamma\cdot p - m)e^{-ip\cdot x}$.
Instead of considering
functions on the manifold $X~=~G/K$, it is preferable to consider 
the action of the Dirac operator on the space of sections of bundles on $X$,
which can also be used to obtain a decomposition of higher-spin fields
analogous to the generalized plane wave expansion of scalar fields.

\vfill
\eject
\centerline{\bf References}

\item{[1]}  G. Dixon, Il Nuovo Cimento ${\underline {105B}}$ (1990) 349 - 364 

\item{[2]}  S. Davis, `Connections and Generalized Gauge Transformations',
University of Cambridge preprint (1996) DAMTP-R/94/49, hep-th/9601178
\item{[3]}  M. de Roo, H. Suelmann and A. Weidmann, Phys. Lett.
${\underline{B280}}$ (1992) 39 - 46 
\hfil\break
H. Suelmann, Int. J. Mod. Physics D, Vol. 3, No. 1 (1994) 285 - 288 
\item{[4]}  N. Ya. Vilenkin, ${\underline {Special~Functions~and~the~
Theory~of~Group}}$
\hfil\break
${\underline{Representations}}$, Translations of Mathematical Monographs,
Vol. 22 (Providence: American Mathematical Society, 1968)

\item{[5]}  S. Helgason, `Lie Groups and Symmetric Spaces' in 
${\underline{Batelle~Rencontres}}$ eds. C. De Witt and J. Wheeler 
(New York: W. A. Benjamin,
1986) 1-71
\hfil\break
S. Helgason, ${\underline {Groups~and~Geometric~Analysis}}$ (New York: Academic
Press, 1984)

\item{[6]}  S. Davis, `Quantum Fields in Anti-de Sitter Space', Knight Prize 
Essay, University of Cambridge (1984)
\hfil\break
S. Davis, ${\underline {Supersymmetry~in~Anti-de~Sitter~Space}}$, 
Ph. D. thesis,
University of Cambridge (1985)

\item{[7]}  S. Helgason, 
${\underline{Geometric~Analysis~on~Symmetric~Spaces}}$, Mathematical Surveys
and Monographs, Vol. 39 (Providence: American Mathematical Society, 1994)

\item{[8]}  D. W. Dusedau and D. Z. Freedman, Phys. Rev. ${\underline {D33}}$
(1986) 389 - 394
\item{[9]}  A. Uhlmann, Czec. J. Phys. ${\underline{29}}$ (1979) 117 - 126

\item{[10]}  L. Castell, Nuovo Cim. A ${\underline{61}}$ (1969) 585 - 592

\item{[11]}  C. P. Burgess and C. A. Lutken, Phys. Lett. ${\underline{B153}}$
(1985) 137 -141

\item{[12]}  P. Breitenlohner and D. Z. Freedman, Ann. Phys.
${\underline{144}}$ (1982) 249 - 281

\item{[13]}  W. Heidenreich, Phys. Lett. ${\underline{B110}}$ (1985) 461 - 464

\item{[14]}  P. Candelas and D. J. Raine, Phys. Rev. ${\underline{D12}}$
(1975) 965 - 974

\item{[15]}  S. J. Avis, C. J. Isham and D. Storey, Phys. Rev. 
${\underline {D18}}$ (1978) 3565 - 3576

\item{[16]}  I. S. Gradshteyn and I. M. Ryzhik, ${\underline{Table~of~
Integrals,~Series~and~Products}}$ (New York: Academic Press, 1980)  

\item{[17]}  A. Kempf, `On Path Integration on Noncommutative Geometries',
University of Cambridge preprint, DAMTP/96-10, hep-th/9603115

\item{[18]}  I. Bars, `Curved space-time strings and black holes', Strings
and Symmetries 1991 (River Edge, NJ: World Scientific Publishing, 1992)
135 - 145
\hfil\break
I. Bars, `Curved space-time geometry for strings and affine non-compact
algebras' ${\underline{Perspective~in~Mathematical~Physics}}$, Conf. Proc.
Lecture Notes Math. Phys. III (Cambridge, MA: International Press, 1994)
51 - 76

\item{[19]}  M. B. Green, J. H. Schwarz and E. Witten, ${\underline{Superstring
~Theory}}$ (Cambridge: Cambridge University Press, 1987)
  
\item{[20]}  T. S. Bunch and L. Parker, Phys. Rev. ${\underline{D20}}$ (1979)
2499 - 2510

\item{[21]}  T. S. Bunch, Gen. Rel. Grav. Vol. 13, No. 7 (1981) 711 - 723 

\item{[22]}  N. Birrell and P. C. W. Davies, ${\underline{Quantum~Fields~in
~Anti-de~Sitter~Space}}$ (Cambridge: CUP, 1982)   

\item{[23]}  C. Fronsdal, Rev. Mod. Phys. ${\underline{37}}$ (1965) 221 - 224
\hfil\break
C. Fronsdal, Phys. Rev. ${\underline{D10}}$ (1974) 589 - 598

\item{[24]}  F. Antonsen and K. Bormann, `Propagators in Curved Space',
hep-th/9608041

\item{[25]}  R. Brunetti and K. Fredenhagen, `Interacting Quantum Fields
in Curved Space: Renormalizability of $\phi^4$', Talk given at Conference
on Operator Algebras and Quantum Field Theory, Rome, Italy, 1 - 6 July 1996,
gr-qc/9701048

\item{[26]}  I. M. Gel'fand, M. I. Graev and N. Ya. Vilenkin,
${\underline{Generalized~Functions}}$, Vol. 5: Integral Geometry and
Representation Theory (New York: Academic Press, 1966)

\end